\documentclass[table]{aa}
\pdfoutput=1
\usepackage[utf8]{inputenc}
\usepackage{amsmath,amssymb,graphicx,xcolor}
\usepackage[]{hyperref}
\usepackage[noabbrev]{cleveref}
\usepackage{multirow}
\usepackage[scr=rsfs]{mathalpha}
\usepackage{graphicx}
\usepackage{subfig} 

\title{The shape of DM halos: a new fundamental cosmological invariance}
\date{\today}

\author{Jean-Michel Alimi \and R\'emy Koskas}

\institute{Laboratoire Univers et Théories (LUTH), CNRS, Observatoire de Paris, PSL Research University\\
5 place Jules Janssen, 92195 Meudon, France\\
\email{jean-michel.alimi@obspm.fr}; \email{remy.koskas@obspm.fr}}

\date{Received X / Accepted Y}

\abstract {In this article, we focus on the complex relationship between the shape of dark matter (DM) halos and the cosmological models underlying their formation. We have used three realistic cosmological models from the DEUS numerical simulation project. These three models have significantly distinct cosmological parameters ($\Omega_m$, $\sigma_8$, and $w$) but their cosmic matter fields beyond the scale of DM halos are quasi-indistinguishable, providing an exemplary framework to examine the cosmological dependence of dark matter halo morphology. 

Firstly, we have developed a robust method for measuring the halo shapes detected in numerical simulations. This method avoids numerical artifacts on dark matter halo shape measurements, induced by the presence of substructures depending on the numerical resolution or by any spherical a priori that does not respect the triaxiality of DM halos. We then obtain a marked dependence of the halo's shape both on their mass and the cosmological model underlying their formation. First, as is well known, the more massive the dark matter halo, the less spherical it is and we find that the higher the $\sigma_8$ of the cosmological model, the more spherical the DM halos. Then, by re-expressing the properties of the shape of the halos in terms of the non-linear fluctuations of the total cosmic matter field or only of the cosmic matter field internal to the halos, we manage to make the cosmological dependence disappear completely. This new fundamental cosmological invariance is a direct consequence of the non-linear dynamics of the cosmic matter field. As the universe evolves, the non-linear fluctuations of the cosmic field increase, driving the dense matter halos towards sphericity. The deviation from sphericity, measured by the prolaticity, triaxiality, and ellipticity of the dark matter halos, is therefore entirely encapsulated in the non-linear power spectrum of the cosmic field.

From this fundamental invariant relation, we can reconstruct with remarkable accuracy the root mean square non-linear fluctuations of the cosmic matter field and, consequently, the power spectrum of the cosmic matter field in which the halos formed. We also re-find the $\sigma_8$ amplitude of the cosmological model that governs the cosmic matter field at the origin of the DM halos. Our results therefore highlight, not only the nuanced relationship between dark matter halo formation and the underlying cosmology but also the potential of dark matter halo shape analysis as a powerful tool for probing the non-linear dynamics of the cosmic matter field.}

\begin{document}
\maketitle 
\nolinenumbers 
\section{Introduction}
\label{introduction}
The formation and evolution of large-scale structures in the universe are among the most fundamental topics in physical cosmology. Dark matter (DM) halos, gravitationally bound concentrations of dark matter, play an essential role in the complex, non-linear processes involved in forming cosmic structures. These halos act as the scaffolding on which visible matter gathers, influencing, as we know, the formation of galaxies, clusters of galaxies, and superclusters of galaxies. Therefore, understanding the properties and behavior of DM halos is essential to unravel the mysteries of cosmic structure formation and their dependence on cosmological models.

Much cosmological research has focused on various properties of DM halos, such as their mass function, internal structure, concentration, or clustering. These studies have shed light on the processes underlying the formation and evolution of DM halos, providing insight into their interactions, mergers, and hierarchical growth (see for example \cite{mo2010} and references therein). In addition, numerous researches have revealed how the properties of DM halos can also depend on the underlying cosmology \citep{1993MNRAS.262.1023W,Tinker2010,Bhattacharya2013,Ludlow2017}.

Among these properties of halos, the shape of DM halos is always an important and intense area of active research interest (see a review in \citet{Limousin2013}). Indeed, the morphology of DM halos provides valuable information on their formation history, the dynamics of their assembly, and their interactions with their environment. Analysis of the shape of DM halos is also crucial for linking theoretical developments to observations \citep{Me2010,Battaglia2012,Lee2018,chira2021}.

It is well-known thanks to simulations and observations that DM halos are not simply spherical objects \citep{Kasun2005, Allgood2006, hayashi2007, Despali2012, Butsky2016, 10.1093/pasj/psw088,Prada2019}. On the contrary, they exhibit triaxiality, their shapes marked by three distinct axes of different lengths. This triaxial nature of DM halos reflects the complex interaction of gravitational forces during their formation. It has implications for their dynamical properties and their interactions with other halos, it influences also the galaxy formation process, and finally, it undoubtedly depends on cosmology.  We especially address this last question in this paper. 

Measuring with high precision the shape of DM halos is no easy task \citep{Zemp2011}. Indeed, such measurements require careful consideration of potential numerical and methodological biases. Such biases can arise from a variety of sources, such as limited numerical resolution, particle sampling effects, halo detection algorithms, or assumptions made in shape measurement techniques. Therefore, it is important to have a reliable method that minimizes possible biases and thus obtain reliable and accurate shape measurements.

In this paper, we first present such a procedure for measuring the shape of DM halos, using reliable techniques that mitigate potential biases. With this procedure, the analysis of DM halos from DEUS numerical simulations \citep{Alimi2010,Rasera2010,Alimi2012,Reverdy2015} first shows a clear dependence of the shape of the halos on their mass and on the cosmology in which they formed. We then show that this cosmological dependence disappears when the shape properties are expressed in terms of the non-linear fluctuations of the matter they contain. 

The universal relation between the shape properties of the halos and the non-linear root-mean-square (rms) fluctuations of the cosmic matter field inside the halos that was highlighted then persists when the non-linear fluctuations are calculated on the total cosmic matter field smoothed on the scale of the halos. This makes it possible to deduce a new universal relation, again independent of cosmology, between the non-linear fluctuations of the matter field internal to the halos and the non-linear fluctuations of the total cosmic matter field. Finally, from the universal relation between halos shape and non-linear fluctuations of cosmic matter field, we reconstruct the non-linear power spectrum at the scale of the halos and also predict the $\sigma_8$ amplitude of the fluctuations of the underlying cosmological model in which the halos were formed.
 
The paper is structured as follows. In section \ref{numerical}, we describe our reliable procedure for measuring the shapes of DM halos where we address the complex challenges posed by limited numerical resolution, substructure identification, and other potential biases. We thus get an accurate and robust way to measure the shape of DM halos. In Section \ref{triaxiality}, we study the cosmological dependence of dark matter halo shapes using data simulations from the DEUS project. We explore how different cosmological models influence the triaxial features of the halos. In section \ref{cosmological}, we highlight the universal relation of the properties of halos shape with the rms fluctuations of the internal matter of the DM halos. We then extend such a fundamental relation to rms non-linear fluctuations of the total cosmic matter field. In section \ref{reconstruction}, we present how to utilize this universal relation between halo shape and non-linear rms fluctuations of cosmic matter field to reconstruct its non-linear power spectrum.  This validates the effectiveness of our method. Finally, in the last section, we conclude and discuss potential limitations and some possible wider implications of our results with possible future directions for potential probing of some aspects of fundamental physics. 

\section{Numerical simulations, cosmological models and DM halos catalog}
\label{numerical}
\subsection{Numerical simulations and cosmological models}
\label{sub-numerical}
\label{secdm}
Over the last few decades, numerical simulations have made it possible to study the formation of cosmic structures in ever larger volumes of the Universe and over ever longer periods of the Universe's history. The study of gravitational collapse during the formation of cosmic structures has thus become increasingly accurate (Numerous reviews on this topic are available, one example being \citet{Kuhlen}). 
In this article, we use numerical data from "Dark Energy Universe Simulation" \footnote{\url{www.deus-consortium.org}}. DEUS simulations \citep{Alimi2010, Rasera2010aip, Alimi2012, Reverdy2015} are high-performance N-body simulations; they reproduce cosmic structure formation assuming various dark energy models. The results presented in this paper have been obtained from three reference main numerical simulations corresponding to three cosmological models: (i) The concordance model $\Lambda $CDM, (ii) Ratra-Peebles quintessence model RPCDM \citep{peebles1988}\footnote{The simulation of the RPCDM cosmological model includes the primordial imprint of the quintessence scalar field on the initial spectrum of perturbations, it is however essentially identifiable with a model with a quasi-constant equation-of-state parameter for dark energy $w_0=-0.8$ \citep{Alimi2010}} and (iii) a phantom model (with a constant equation of state $w_0<-1$) that we denote $w$CDM. The cosmological parameters of each of these models have been chosen in agreement with both SNIa and CMB WMAP7 constraints \citep{komatsu2011}, as summarized in table \ref{param}. Such models are thus said to be \textit{realistic} \citep{Alimi2010} and their present cosmic fields are naturally very close to each other. These simulations were performed in a $L=648$ \textrm{Mpc}/h periodic box using $N=2048^3$ particles. The analysis of possible numerical effects on the measures of shape parameters of DM halos, discussed in section \ref{triaxiality} has been performed with additional numerical data from DEUS simulations with different mass and space resolutions i.e. with different numbers of particles and different computing box sizes. 

\begin{table}
    
    \centering
    \begin{tabular}{|c||c|c|c|}
    \hline 
    Model & $\Lambda $CDM & RPCDM  & $w $CDM \\ 
    \hline
    \hline
    $\Omega_m$  &0.2573 &0.23 & 0.275 \\
    \hline
    Particle mass ($M_{\odot}/h$) & $2.3 \cdot 10^{9}$ & $2.0\cdot 10^{9}$ & $2.4\cdot 10^{9}$ \\
    \hline
    $\sigma_8^{Gauss}$ & 0.83 & 0.68& 0.88\\
    \hline
    $\sigma_8^{Bessel}$ & 0.8& 0.66 &0.85 \\
    \hline
    $w$ & -1.0& -0.8& -1.2 \\
    \hline
    $\delta_c$&1.673&1.672&1.674\\
    \hline
    \end{tabular}
    \caption{Parameters of Dark Energy Simulations. For all of those we used, $\Omega_K=0$, $n_s=0.9630$, $h=0.72$, $L=648$ \textrm{Mpc}/h and $N=2048^3$ particles. In addition, for initial power spectrum computation $\Omega_b=0.04356$ (and baryons are then supposed to follow dark matter dynamics). The $\sigma_8$'s are the rms fluctuations at 8 \textrm{Mpc}/h, computed on the power spectrum \textit{linearly} extrapolated to $z=0$, and using a Gaussian $W(x)=\exp -\frac{x^2}{10}$ or a Bessel $ W(x)=3\frac{\sin(x)-x \cos(x)}{x^3}$ window function. $\delta_c$ is the critical spherical overdensity at $z=0$   }\label{param}
\end{table}

\begin{figure}[h!]
    \includegraphics[width=8cm]{"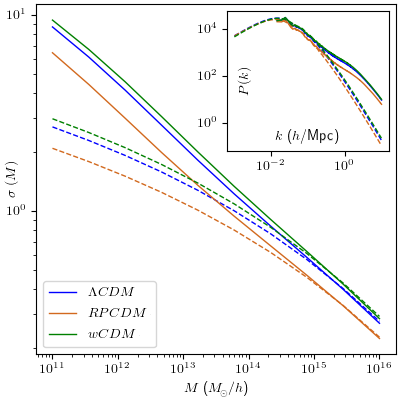"}
    \caption{The root-mean-square (rms) fluctuations as a function of the smoothing mass $M$. The RMS $\sigma_L$ of the \textit{linear} power spectrum  (calculated by CAMB) are dotted lines. The RMS $\sigma_{NL}$ of the \textit{non-linear} power spectrum (calculated from $z=0$ snapshots of DEUS simulations) are in solid lines. $\sigma_{NL}(M)$ behaves essentially as a power law whose slope and intercept depend on the cosmology. The inner panel contains the corresponding linear and non-linear power spectra, $P_L$ and $P_{NL}$. At high mass (or low $k$), the linear and non-linear rms fluctuations (as well as the power spectra) are superimposed, as expected.}
    \label{rms-pk}
\end{figure}

\begin{figure}[h!]
    \includegraphics[width=8cm]{"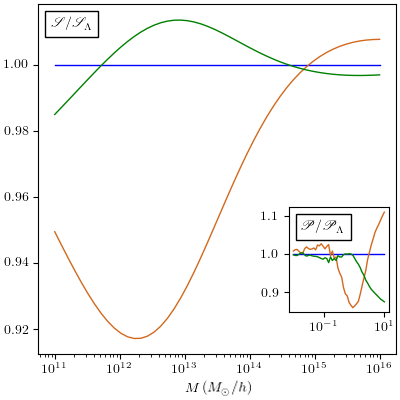"}
    \caption{The ratio to the $\Lambda$CDM (blue) model as a function of the smoothing mass $M$ of the purely non-linear rms fluctuations $\mathscr{S}={\sigma_{NL}}/{\sigma_L}$ for RPCDM (orange) and $w$CDM (green) models. In the internal panel, we plot, in the same way, the ratio to the $\Lambda$CDM model of the purely non-linear power spectrum $\mathscr{P}={P_{NL}}/{P_L}$ of each model. We note that $\mathscr{P}_{RPCDM}$ (orange) closely follows $\mathscr{P}_{\Lambda CDM}$ (blue) up to $k_C=0.4\,\,h \, \textrm{Mpc}$, after which it deviates significantly. For the $w$CDM model (green), the purely non-linear spectra are very distinct above $k_C=1.\,\,h\; \textrm{Mpc}$. On the purely non-linear ratios of rms fluctuations, the difference is strong on mass scales $10^{12}$-$10^{14}$ $M_{\odot}/h$. It is precisely this range of mass that we will study from the DM halos} 
    \label{ratiorms}
\end{figure}

To assess the impact of non-linear effects on cosmic structure formation for these models, we first computed at $z=0$ both the linear and non-linear power spectra of the total matter cosmic field. The first one denoted $P_L(k)$, is simply the linear evolution of the initial power spectrum until $z=0$, as given by CAMB \citep{camb}. The non-linear power spectrum $P_{NL}(k)$ is computed from the $z=0$ density field of the numerical simulations' data.  Both are plotted in the internal panel of \cref{rms-pk}. As expected, they differ at small scales ($k>0.2 \,h\textrm{Mpc}$) and merge on larger ones. The same observation holds when smoothing power spectra over mass scales, defining the rms fluctuations: 

\begin{equation}
    \sigma^2_L(M)=\frac{1}{2\pi^2}\int_{0}^{+\infty}k^2 P_L(k) W^2\left[k\cdot \left(\frac{3M}{4\pi\Omega_m\rho_c}\right)^{1/3}\right]\;\mathrm{d}k
\end{equation}

and similarly with $P_{NL}(k)$ for $\sigma^2_{NL}(M)$. We have chosen a Gaussian\footnote{There are alternatives, for example, the Bessel Window $$W(x)=3\frac{\sin(x)-x \cos(x)}{x^3}$$ which is the Fourier transform of a spherical top-hat in real space.} smoothing window $W(x)=\exp -\frac{x^2}{10}$. The $\frac{1}{10}$ factor within our Gaussian window is chosen so that (squared) Gaussian and Bessel windows coincide up to the second order around $0$. Comparing $\sigma_L$ and ${\sigma}_{NL}$, as plotted in the left panel of \cref{rms-pk}, we again notice that the curves superpose on large scales ($M>10^{15} M_{\odot}/h$) and already differ for halos with mass ($10^{14} M_{\odot}/h$). We then defined the \textit{pure} non-linear power spectrum as the ratio between non-linear and linear power spectra, for each cosmological model, $\mathscr{P}(k)=P_{NL}(k)/P_L(k)$ \citep{Alimi2010}. From such a quantity, we can interestingly see that cosmological background leaves an imprint during the differentiation between linear and non-linear scales: the internal panel of \cref{ratiorms} features the pure non-linear power spectrum normalized to the $\Lambda $CDM one, $\mathscr{P}^X(k)/\mathscr{P}^{\Lambda CDM}(k)$. The behaviors of $\mathscr{P}^{RPCDM}/\mathscr{P}^{\Lambda CDM}$ and of $\mathscr{P}^{wCDM}/\mathscr{P}^{\Lambda CDM}$ manifestly differ. The former increases for $k > 0.1 \,h \, \textrm{Mpc}$ while the latter is still decreasing for those modes. Furthermore, the critical mode $k_c$, on which the pure non-linear power spectrum deviates by more than five percent from the $\Lambda $CDM model, heavily depends on the cosmological model: for RPCDM, $k_c \approx 0.3 \,h \, \textrm{Mpc}$ and for $w$CDM, $k_c \approx 2 \,h\, \textrm{Mpc}$. We thus see non-linear evolution does not forget the background cosmology or, in other terms, the coupling modes during non-linear evolution explicitly depend on cosmology. 

Similar remarks can be made about the dependence of rms fluctuations as a function of mass. The mass range spanned by purely non-linear collapse can then be determined by studying the following quantity $\mathscr{S}(M)=\sigma_{NL}(M)/\sigma_L(M)$. The ratio of this quantity to $\mathscr{S}^{\Lambda CDM}$ is plotted as a function of mass in \cref{ratiorms}. $\mathscr{S}^{RP CDM}$ and $\mathscr{S}^{wCDM}$ deviate from $\mathscr{S}^{\Lambda CDM}$ for masses below $10^{14} M_{\odot}/h$. Halos with such masses are therefore especially suited probes for assessing the cosmological footprint on non-linear dynamics, or in other words, it's by studying the properties of halos with masses below $10^{14} M_{\odot}/h$ that we can probe the non-linear power spectrum and its cosmological dependence. In the following, we will study the shape of these halos, and show that when such a feature of halos is carefully evaluated, it effectively constitutes a powerful probe of cosmology. 

It should also be noted that even if the cosmological models we have considered are realistic, this does not mean that the cosmological parameters that define them are not significantly different (See Table \ref{param}). For instance, $\sigma_8$ is about $30\%$ lower in the Ratra-Peebles model than in the phantom model. Similarly, the $\Omega_m$ values vary by about $20\%$ between the RPCDM and $w$CDM models. These differences are much larger than those of the models used by \cite{Despali2014}, where these authors also studied the cosmological dependence of the shape of DM halos, but for cosmological models with very similar cosmological parameters.

\subsection{DM halos in numerical simulations}
\label{dark}

\begin{table*}
    \centering
    \begin{tabular}{|c||c|c|c|}
    \hline
    Model & $\Lambda $CDM & RPCDM  & $w $CDM \\ 
    \hline
    \hline
    Number of halos   $M_{FoF}\in [2.4\cdot 10^{12},1\cdot 10^{14}]$ $M_\odot/h$                  & 411 514 & 338 584 & 438 456 \\
    \hline
    After eliminating crossing-border halos              & 409 682 & 337 031 & 436 552 \\
    \hline
    After eliminating the halos s.t.   $M_{substructures}/M_{FoF}>0.1$         & 368 613 & 297 183 & 395 962 \\
    \hline
    After eliminating the halos s.t.   $M_{200}/M_{FoF}<0.8$ (Selected halos) & 349 914 & 271 764 & 378 887 \\
    \hline
    Proportion of selected halos                         & 85\%    & 80\%    & 86\%   \\
    \hline 
    $\Bar{M}$                 &$9.0 \cdot 10^{12}$  &$8.1 \cdot 10^{12}$   &$9.4 \cdot 10^{12}$\\
    \hline
    Median ${M}$                 &$4.8 \cdot 10^{12}$  &$4.5 \cdot 10^{12}$   &$4.9 \cdot 10^{12}$\\
    \hline
    \end{tabular}
    \caption{Number of DM halos in each simulation: First we count all the halos detected by the FoF algorithm in each cosmological model. We then discarded halos that “crossed” the boundary of the periodic simulation box. We then discard halos where $M_{200}$ is more than $20\%$ different from $M_FoF$. Finally, we discard halos where the mass of the substructures represents more than $10\%$ of the total mass. This gives us the catalogs on which the statistical analyses presented in this paper are based }
    \label{cathalos}
\end{table*}

DM halos were detected in simulations with the Parallel Friends of Friends algorithm with a percolation parameter $b=0.2$. We selected well-resolved halos containing more than $1000$ particles. With the $\Omega_m$ values of the DEUS simulations used, this amounts to selecting halos from the size of a group of galaxies to a cluster of galaxies, with masses between $2.4\cdot 10^{12}$ and $1\cdot 10^{14}$ solar masses (per $h$). We have more than $300,000$ halos for each cosmology. In \cref{cathalos}, we present some characteristics of the halos catalog deduced from each simulation. The halos in these catalogs are particularly suitable for studying the cosmological dependence of non-linear dynamics, as we saw in the previous section.

The few halos that “cross” the boundary of the periodic simulation box are ignored. We also eliminate the objects detected by the FoF algorithm which are only the combination of two clusters of comparable mass, linked by a thin filament. Such bias called bridging \citep{Davis1985}, is a very specific feature of the FoF detection method, so we choose to exclude such objects from our analysis. In practice, we exclude FoF objects whose spherical overdensity mass $M_{200}$ is more than $20\%$ different from the total FoF mass $M_{FoF}$\footnote{FoF mass is the mass deduced by counting all particles included in halos detected by FoF algorithm} or whose substructures (see below) excess $10\%$ of the total FoF mass (because such voluminous substructures are probably neighboring halo falling on the main one). Any detected halo-like structure that presents at least one of these two characteristics has been simply discarded from the sample of halos used in our analysis. 

The halo catalogs constructed for the three cosmological models are presented in Table \ref{cathalos}. They contain a large population of halos (more than 300,000). The mass range of the halos in each catalog perfectly covers the scales on which the strongest cosmological imprints on non-linear dynamics are expected. 
\\
\begin{figure*}
    \subfloat{\includegraphics[width=6cm]{"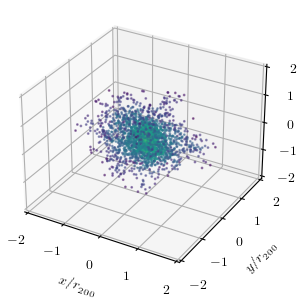"}}
    \subfloat{\includegraphics[width=6cm]{"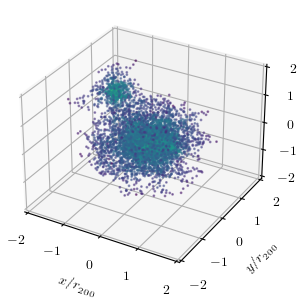"}}
    \subfloat{\includegraphics[width=6cm]{"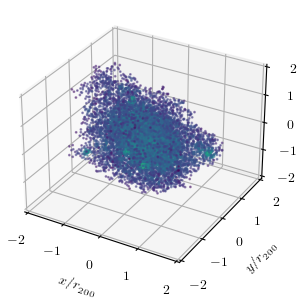"}}
    \\
    \subfloat{\includegraphics[width=6cm]{"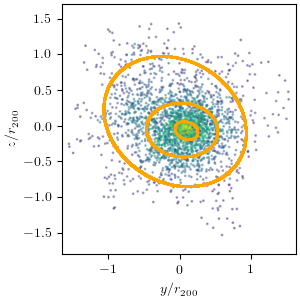"}}
    \subfloat{\includegraphics[width=6cm]{"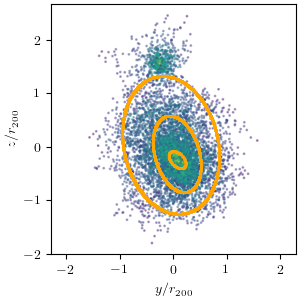"}}
    \subfloat{\includegraphics[width=6cm]{"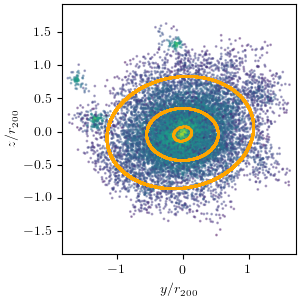"}}
    \\
    \subfloat{\includegraphics[width=6cm]{"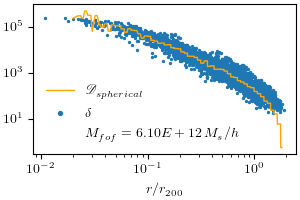"}}
    \subfloat{\includegraphics[width=6cm]{"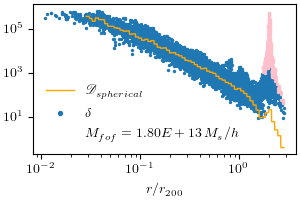"}}
    \subfloat{\includegraphics[width=6cm]{"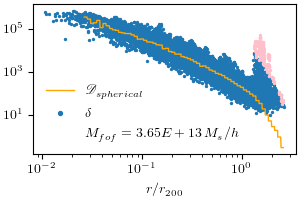"}}
    \caption{Three DM halos with three various numbers of particles (the particle mass is equal to 
    $2.1 \cdot 10^{9} M_{\odot}/h$) in $\Lambda$CDM Dark Energy Simulations. Their mass is respectively from left to right $6.1\cdot10^{12}, 1.8\cdot10^{13}, 3.65\cdot10^{13}$ $M_{\odot}/h$. The first line of this figure shows the three-dimensional representations of the halos: each halo contains numerous substructures, mainly located at the periphery; their number increases with the mass of the halo. The second line shows the ellipsoids corresponding at best to the $\delta=200$, $\delta=2000$ and $\delta=20000$ iso-densities, superimposed on the particles in the $(y,z)$ plane. The graphs in the third row show the point clouds in the $(r,\delta)$ plane, where each point represents a particle located at a distance $r$ from the center of mass of the halo and $\delta$ is the local over-density at the point where the particle is located. The pink dots correspond to particles belonging to substructures (see subsection 2.3). In orange, we plot the average overdensity of the spherical shell at radius $r$. The local density is not a one-to-one function of the distance to the center of mass. This again illustrates the asphericity of the halo. The substructures appear in this representation as a set of particles grouped in the form of local density peaks that are superimposed on the spherical density profiles.}  
    \label{halo3d}
\end{figure*}

\section{Triaxiality of DM halos: Methodological aspects}
\label{triaxiality}
\subsection{Definition of shape parameters}
\label{definition}
While it is usual to model halos, to a first approximation, as a stack of concentric isotropic spheres, we are nevertheless led to enrich their description by taking into account their triaxiality. This triaxiality manifests itself at different levels. In parallel with theoretical knowledge and the numerous observational results, N-body simulations largely confirm that the current shape of dark matter (DM) halos deviates from sphericity: \cite{Jing2002} had already noticed that surfaces of constant local density are well fitted by ellipsoids. Numerous other studies have also concluded that there is significant triaxiality in the shape of the halos \citep{Bailin2005,Kasun2005,  Allgood2006, hayashi2007, Veraciro2011,  Bonamigo2015, Butsky2016, Prada2019}. As a first illustration, we consider the three-dimensional representation of three simulated halos in the $\Lambda $CDM cosmological model, with respective mass $M_{FoF}=6.1\cdot 10^{12} M_{\odot}/h$, $M_{FoF}=1.8\cdot 10^{13} M_{\odot}/h$ and $M_{FoF}=3.65\cdot 10^{13} M_{\odot}/h$ in \cref{halo3d}. The first is rather spherical, and the others, besides the numerous peripheral substructures, are more compressed along one axis. We made it more visible in the second row of this figure where the halos are projected onto the simulation plane $y-z$. The second halo in the central position presents a sub-halo, very off-center and apparently with a non-negligible mass. Finally, in the third row of \cref{halo3d}, we compute the scatter plot in which each point is a particle of the considered halo, its abscissa is its distance to the halo densest point,  its ordinates, in blue, are the local overdensity ($\delta$) where the particle stands. We represent in orange, the overdensity of the spherical shell this particle belongs to (among 64 other, logarithmically spaced).  The fact that there is no one-to-one correspondence between the radius and local overdensity (the cloud is thick) is another illustration of the halo triaxiality so a spherical description is not relevant. We will thus approximate the overall shape of halos as an ellipsoid. However, for a given halo, determining the parameters of the \textit{best fitting} ellipsoid (i.e. size and orientation of its principal axis), must once again be carried out with care. 

The majority of available methods involve the computation of the inertia tensor of the considered halo: 
\begin{equation}
{\mathcal{M}^{ij}_{\mathrm{sample}}=\langle x_ix_j\rangle_{\mathrm{sample}} -\langle x_i\rangle_{\mathrm{sample}}\langle x_j\rangle_{\mathrm{sample}} \;\;\; 1\leq i,j\leq 3}
\end{equation}
where the averages are taken from a certain sample of particles. The choice of the sample of particles to be used in each halo has to be done with delicate. Indeed, taking into account all the particles in a given halo to extract the main ellipsoid is both sensitive to the algorithm used to detect that halo, and to the resolution in which the halo has been simulated. This point will be discussed in more detail in the next section, but in any case, such potential numerical effects need a reliable method for capturing the shape of halos (see section \ref{measure}).

Let us denote $a^2 \geq b^2 \geq c^2$ the eigenvalues of $\mathcal{M}_{\mathrm{sample}}$. One can form the following dimensionless quantities: 
\begin{itemize}
    \item the ellipticity $E=\frac{a-c}{2(a+b+c)}$
    \item the prolaticity $p=\frac{a-2b+c}{2(a+b+c)}$
    \item the triaxiality 
    $T=\frac{a^2-b^2}{a^2-c^2}$
\end{itemize}
The ellipticity quantifies the deviation from sphericity while the two others allow distinguishing two kinds of flattening: an oblate $(p\rightarrow -\infty, T\rightarrow 0)$ ellipsoid looks like a pancake while the shape of a prolate ellipsoid $(p\rightarrow +\infty, T\rightarrow 1)$ is close to a filament. For a perfect sphere, $p$ vanishes and $T$ is not defined. In practice, the calculation of these dimensionless shape parameters for each halo will thus consist of determining the sample of particles that will be taken into account to calculate the inertia tensor, among all the particles that make up this halo. 

\subsection{Measure of Shape parameters}
\label{measure}
\subsubsection{Substructures removal}
\label{substructures}
\begin{figure}[h!]    
\includegraphics[width=8cm]{"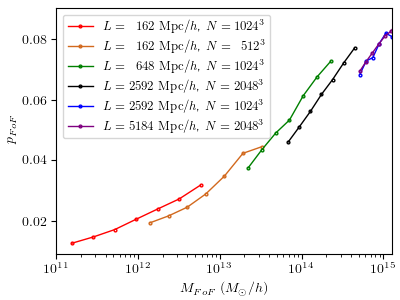"}
    \caption{Dependence of the shape of the FoF halos according to the numerical resolution of the simulations in which they were formed. All the simulations correspond to the same $\Lambda$CDM model. Each simulation was performed with a specific computing box size $L$ and contains a specific number $N$ of particles. Consequently, the particle masses $m_p$ are specific to each simulation. We detected the FoF halos in each of the simulations. We computed their FoF mass, and (without further processing) their shape parameters. We plot the median prolaticity of the halos in each mass range.  We first notice that the shape of the FoF halos does indeed depend on the numerical resolution. Halos of $10^{12}-10^{13}\, M_{\odot}/h$ are more prolate in the simulation $L=162\,\textrm{Mpc}/h$ with $N=1024^3$ particles than in the less resolved simulation $L=162\,\textrm{Mpc}/h$ with $N=512^3$ particles. We also note that this effect depends only on the ratio $N/L^3$ (that is $m_p$). The halos corresponding to the blue and purple curves developed in boxes with different $L$ and $N$, but with the same mass of particles, have not a notably distinct median prolaticity.}      
    \label{multFoF}
\end{figure}

First of all, we need to deal with the substructures that can make a significant contribution to the mass and shape of the halos, but whose presence depends on the numerical resolution. We have calculated the shape parameters of simulated halos in boxes of different sizes $L$ and with different numbers of particles $N$. We show in the following the results for prolaticity $p$ for the cosmological model $\Lambda$CDM using DEUS data with cosmological parameters $\Omega_m=0.25$, $h=0.72$ and $\sigma_8=0.81$. The conclusions are unchanged with the cosmological models, RPCDM and $w$CDM. If the inertia tensor is calculated from all the particles of the FoF halo, its eigenvalues will depend strongly on the numerical resolution due to the presence of the substructures. For each of these simulations with different numerical resolutions for the same cosmological model, we calculated the median of the prolaticity $p$ of all the halos in each mass range. These results are presented in Figure \ref{multFoF}. Several mass domains, determined by the resolution $N/L^3$ (the latter being proportional to the mass of the particle, $m_p$) are thus probed. We observe that simulations with identical resolutions, such as $(L, N)=(2592,1024^3)$ and $(5184,2048^3)$ produce halos that, for the same total mass, have the same shape (in this case, the same prolaticity). However, for a given halo mass, the higher the resolution of the simulation (the lower $m_p$), the more prolate the halos: for example, the $10^{13}$ $M_{\odot}/h$ halos are 15 \% more prolate in the $(L, N)=(2592,2048^3)$ simulation than in the $(L, N)=(648,1024^3)$ simulation. This effect is linked to the presence of the substructures. The inertia tensor is mainly sensitive to particles located at the edge of the halo since their contribution is equal to the square of their distance from the center of the halo. Since substructures are, by nature, objects that are not well mixed with the rest of the halo, many of them still fall and tend to be located at the edge of the halo. At lower resolutions, fewer substructures form and those that do contain a smaller number of particles. The edge of the halo and its shape are therefore modified as a function of the presence and mass of the substructures. As the numerical resolution in mass decreases, the halos formed are measured as necessarily more isotropic and their prolaticity is lower. 

To better test such a hypothesis, we erased the substructures in the halos and observed how this affects their shape as a function of resolution. Substructures are mainly abnormally dense regions in a given halo. To locate them (and thus be able to erase them), halos could be detected using the FoF algorithm but with a lower percolation parameter (see \cite{Jing2002}).  However, whereas  \cite{Jing2002} uses extremely well-resolved halos for his analysis (more than $10^6$ particles per halo), we have "only" a few thousand particles per halo (obviously we prefer to use the number of halos available for our analysis, that is several hundred thousand, unlike \cite{Jing2002} which has only a few halos). Consequently, we here prefer another procedure in which we will compare the density at any point where a particle is located with the density of the spherical shell to which it belongs. The procedure for erasing the substructures is then divided into four stages: (i) We calculate the density $\delta$ at any point where the particles are located. To do this we use an SPH kernel \citep{gingold1977,lucy1977} including 32 neighbors\footnote{They are contained in a sphere of radius $r_{32}$. The smoothing radius is then $h=r_{32}/2$.} $\delta$ is calculated in units of $0.25\rho_c$ (and not $\Omega_m\rho_c$) to adopt an agnostic point of view on the exact value of $\Omega_m$, (ii) We define the center of the halo as the position of the particle with the highest $\delta$. This center can generally be different from the center of mass of the halo. From now on, it will be the origin of the coordinates, (iii) The halo is then divided into 64 concentric spherical shells; thus, for a particle at position $\mathbf{r}^{(i)}$, we note $\mathscr{D}_{\mathrm{spherical}}(r_i)$, the overdensity (wrt $0.25\rho_c$) of the shell to which this particle belongs and (iv) Finally, in the point cloud $(r^{(i)},\delta^{(i)})$ (third line of fig \ref{halo3d} for example), the substructures correspond to the peaks. Hence, the following definition of substructures: these are particles such that $\delta^{(i)}>\eta\cdot\mathscr{D}_{\mathrm{spherical}}(r^{(i)})$. This is because the overdensity of a spherical shell is averaged over the angles and is therefore much less sensitive to the presence of substructures. The parameter $\eta$, which quantifies the rigor of substructure elimination, has yet to be determined. There is a variant in which the local density is compared with the mean local density and the standard deviation in the shell, see \citet{Stapelberg2020}. Note again that the fact that the point cloud $r,\delta$ does not represent a single-valued function (the curve $\delta(r)$ is thick) is a clear manifestation of the non-sphericity of the halo.   

\begin{figure*}
    \includegraphics[width=17cm]{"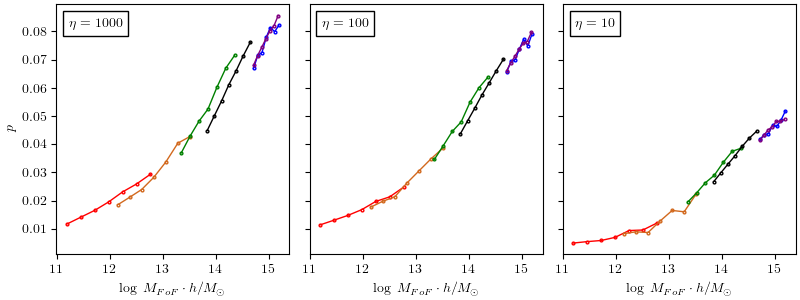"}
    \caption{Removal of substructures and resolution effect. For the $\Lambda$CDM model, we consider several simulations with different computational box sizes and different numbers of particles. In each FoF halo, we remove substructures at a certain $\eta$ level (the larger the $\eta$, the smaller the number of particles designated as belonging to the substructures — see main text). The prolaticity is then calculated on these treated halos where the substructures have been removed. It is plotted as a function of halo mass. For $\eta=1000$ (left), the resolution effect mentioned above is still present: in a given mass range, increasing the mass resolution increases the median prolaticity of the treated halos. This effect is strongly attenuated when we choose $\eta\leq 100$ (middle): the mass-prolaticity curves corresponding to the different resolutions are superimposed in the common halo mass domains.  This allows us to attribute the numerical resolution dependence of the FoF halo shapes to the presence of substructures. However, at $\eta=10$ (right), the overall prolaticity collapses: a too large fraction of the particles in the halo were considered as belonging to substructures and were removed. The ellipsoidal nature of the resulting halos is then strongly altered, and they are about half as prolate as they were before the substructures were removed. A reliable level, independent of the numerical resolution of the substructure suppression, can be considered when $\eta=100$.}
    \label{figshapeta}
\end{figure*}

\begin{figure}[h!]
    \includegraphics[width=8cm]{"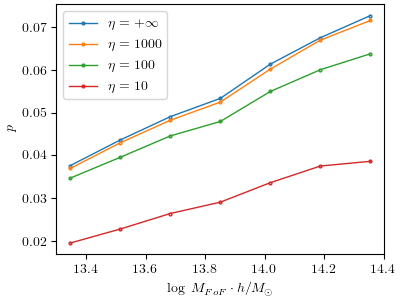"}
    \caption{Effect of the stringency of substructure removal according to the $\eta$ parameter on the overall prolaticity of halos. In the simulation $L=648\;\textrm{Mpc}/h, N=1024^3$, we plot the median mass-prolaticity curves for halos treated with different substructure removal parameters $\eta$. By choosing $\eta=+\infty$, no substructures are removed. The corresponding curve is only about $10\%$ larger than that of the halo shapes at $\eta=$100, whereas it is twice as large as the halo prolaticity treated with $\eta=$10. The treatment of halos with $\eta=100$ thus seems to be the most reasonable and does not artificially reduce their shape measurements to that of a sphere.}
    \label{eta}
\end{figure}

We then calculated and diagonalized the mass tensor considering only particles that do not belong to substructures, i.e. all particles such that $\delta\leq \eta \mathscr{D}$, with respectively $\eta=1000,100,10$. The median prolaticity of the halos (with or without substructures) is plotted as a function of the total FoF mass of each halo. The results are shown in figure \ref{figshapeta}: if the aforementioned resolution effect seems to be tempered with moderate ($\eta=100$) and strict ($\eta=10$) removal of substructures, it survives when removing only a few particles with $\eta=1000$, hence confirming our assumption. However, a too-aggressive suppression of substructures with $\eta=10$ tends to alter the ellipsoidal characteristic of the halo itself, with many particles satisfying $\frac{\delta^{(i)}}{\mathscr{D}_{\mathrm{spherical}}(r^{(i)})}\geq 10$, without belonging to substructures. Their presence, and more generally the thickness of the point cloud $(r,\delta)$, is then simply a manifestation of the non-sphericity of the halo, as mentioned above. In figure \ref{eta}, we see the median prolaticity for different $\eta$, in simulations with $648 \textrm{Mpc}/h$ computing box and $1024^3$ particles. The prolaticity of halos with $\eta=100$ is only $10\%$ different from the prolaticity deduced from the full FoF halo, however, it reduces to half the original prolaticity when $\eta=10$ - a reasonable definition of substructures should not lead to such a drastic sphericization of the halo when they are removed. As a conclusion, we will consider halos without the so-called “$\eta=100$” substructures (i.e., without the particles such that ${\frac{\delta^{(i)}}{\mathscr{D}_{\mathrm{spherical}}(r^{(i)})}\geq 100}$). This is a good compromise between the need to attenuate resolution effects on the one hand and to preserve the shape of the halo on the other hand. 
     
\subsubsection{Isodensity selection} 
\label{isodensity}

The inertia tensor is dominated by particles located near the halo's edge, which is precisely sensitive to the halo detection method.  In particular, while the FoF algorithm applies without any a priori on the shape of the halos, this is not the case for the SOD algorithm which, by construction or by definition, favors significantly more spherical halos, which ultimately biases the measurements of the shape distributions of the detected halos. More precisely still, to obtain an estimate of the halo shape that is independent of the detection algorithm, we need to attenuate the contribution of particles located very close to the edge of the halo. 

To achieve this, some authors have proposed normalizing the contribution of particles to the inertia tensor by the square of their distance from the center \citep{Bailin2005}, but in this case the inertia tensor normalized in this way no longer has the physical dimension of a classical inertia tensor. Other authors \citep{Allgood2006, Despali2012} have developed shape calculation methods with multiple iterations during the diagonalization of the inertia tensor until these successive iterative measurements converge.  \citet{Zemp2011} show that this method favors the contribution of particles very internal to the halo, which can again bias the calculation of the overall shape of the halo. We finally preferred a third method, used in particular by \citet{Jing2002}. This one begins by calculating an initial estimate of the inertia tensor from the particles included in the shell around a chosen iso-density $\delta=\delta_0$ of the halo, that is, in practice, the set of particles for which $0.8\delta_0\leq \delta \leq 1.2\delta_0$. The connectedness of surfaces of a given density is guaranteed by the prior elimination of substructures. All the particles belonging to this thick shell contribute to the corresponding mass tensor, $\mathcal{M}_{\mathcal{S}_0}$. By diagonalizing it, we obtain an ellipsoidal fit of this shell. The lengths of the semi-axes of this ellipsoidal shell are $a_{\delta_0}=\sqrt{3\lambda_1},b_{\delta_0}=\sqrt{3\lambda_2}, c_{\delta_0}=\sqrt{3\lambda_3}$ where $\lambda_1\geq\lambda_2\geq\lambda_3$ are the eigenvalues of $\mathcal{M}_{\mathcal{S}_0}$. Then, to reduce any noise, we add an integration step where we consider all the particles in the halo that are \textit{interior} to the ellipsoidal shell (defined by $a_{\delta_0},b_{\delta_0},c_{\delta_0}$).  All these particles fill the ellipsoidal volume (and not just an ellipsoidal \textit{shell}). To obtain the best-fitting parameters, we then calculate the mass tensor $\mathcal{M}_{\mathcal{E}_0}$ from the positions of the latter particles. By diagonalizing $\mathcal{M}_{\mathcal{E}_0}$, we obtain the new (and final) measure of the lengths of the half-axes of the halo: $a_{\delta_0}=\sqrt{5\mu_1},b_{\delta_0}=\sqrt{5\mu_2},c_{\delta_0}=\sqrt{5\mu_3}$ where $\mu_1\geq \mu_2\geq\mu_3$ are the eigenvalues of $\mathcal{M}_{\mathcal{E}_0}$.

The last values obtained of $a,b,c$ are only slightly different from the previous ones (which were calculated using the $\mathcal{M}_{\mathcal{S}_0}$ eigenvalues), as shown in \citet{Zemp2011}. However, by proceeding in this way, we are assured that the estimate of the halo half-axis lengths will be less sensitive to details of the very outer shape of the halo and in particular to particles that belong to the (thick) isodensity $\delta=\delta_0$, but are located far from the corresponding ellipsoidal fit. Finally, this method is relatively insensitive to the halo detection algorithm (FoF or SOD) used, as long as the edge of the halo does not intersect the chosen isodensity. 

\begin{figure}[h!]
    \includegraphics[width=8cm]{"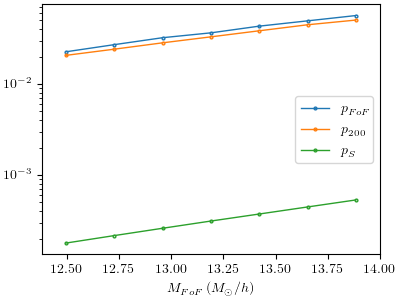"}
    \caption{Effect of spherical prior on shape measures. We consider FoF halos of $\Lambda $CDM Universe simulated in a $L=648 \textrm{Mpc}/h, N=2048^3$ cube. Their prolaticity is computed: by diagonalizing the mass tensor of halos treated in three ways:   through the procedure described in the text (substructures removal and isodensity selection) – this is the $p_{200}$ curve (orange); by simply ignoring particles out of the sphere whose spherical overdensity is $\Delta=200$. This boils down to introducing a spherical prior, and the corresponding curve is denoted $p_S$ (green); or with no treatment on FoF halos - $p_{FoF}$ (blue). It turns out that computing shapes of halos truncated by spheres considerably bias the measure of prolaticity (and of the other shape parameters), artificially bringing it closer to $p=0$.}
\label{peps}
\end{figure}

In \cref{peps}, we represent the median prolaticity of the $\Lambda$CDM halos, calculated in three different ways: (i) the blue curve is obtained from all the particles in the FoF halo, (ii) from all the particles in the FoF halo, we calculate the mass tensor, keeping only the particles belonging to the \textit{mean spherical density} ball defined by $\Delta=200$, the resulting prolaticity is the green curve and is about 100 times weaker than the previous estimate (blue curve). In other words, forcing the halo's edge by identifying it as a sphere, even though the particles are not uniformly distributed in it, induces a spherical bias in the measurement of the shape which artificially tends $p$ to approach $0$. This invalidates (de facto) such an approach. Finally (iii) among all the particles in the FoF halo, we select the particles after extraction of the substructures and according to the $\delta=200$ isodensity mode as explained above. We then obtain the orange curve, which provides a very good measure of the shape of the overall halo, insensitive to resolution, which is ultimately quite close to but differs precisely from the estimate of the shape deduced from all the particles in the FoF halo.

\section{Cosmological origin of the DM halos shape}
\label{cosmological}
\subsection{Cosmological imprints on the DM halos shapes}
\label{cosmologicalimprints}
\begin{figure}[h!]
    \includegraphics[width=8cm]{"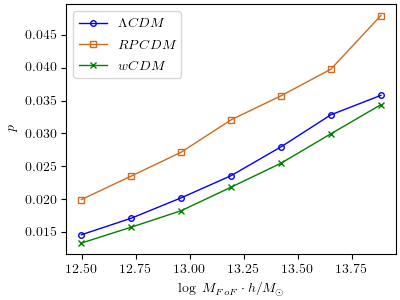"}
    \caption{Median prolaticity as a function of halo mass for each tested cosmological model. The more massive are the halos, the less they collapse and the higher their prolaticity. One notices that both the slope and intercept of the curves depend on the cosmological model: RPCDM halos are more prolate than the $\Lambda$CDM model. This flows from the fact that $\sigma_8$ is lower in the chosen Ratra-Peebles cosmology than in the $\Lambda$CDM one; the halos of the first have thus collapsed more recently and are therefore less spherical. }
    \label{3}
\end{figure}

\begin{figure*}
    \subfloat[Median middle-to-major axis ratio versus halo mass]{\includegraphics[width=8cm]{"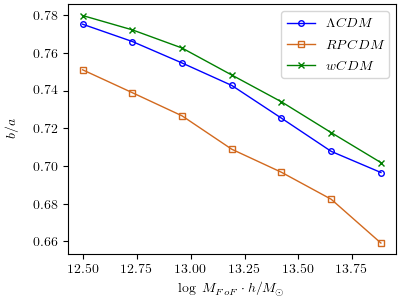"}}
    \subfloat[Median triaxiality versus halo mass]{\includegraphics[width=8cm]{"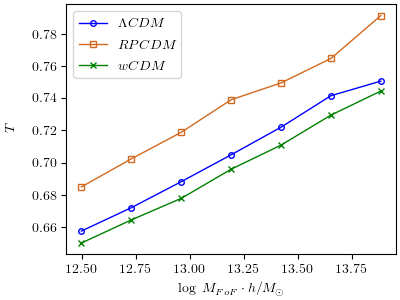"}}\\
    \subfloat[Median minor-to-major axis ratio versus halo mass]{\includegraphics[width=8cm]{"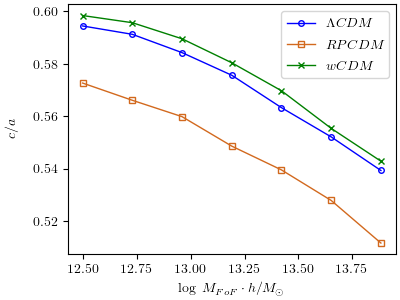"}}
     \subfloat[Median ellipticity versus halo mass]{\includegraphics[width=8cm]{"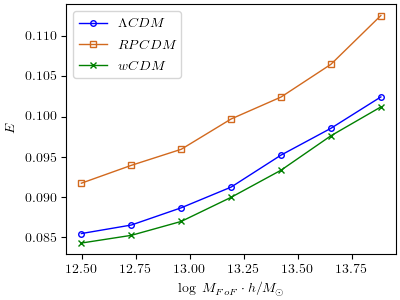"}}
     \caption{Dependence of other halos shape parameters on cosmology and halo mass. The halos of cosmological models with a lower $\sigma_8$ that have collapsed more recently are less spherical.}
     \label{4}
\end{figure*}

We plot in \cref{3,4} the median value in each mass range of the dimensionless shape parameters ($p$, $T$, $E$, axis ratio) as a function of their mass, for the three cosmological models. These indicators were calculated using the procedure presented previously, which ensures that our measurements are robust to the numerical resolution and the halo detection algorithm used while respecting their triaxialities. They are represented as a function of the total mass (FoF) of the halo\footnote{The mass contained in the region delimited by the shell $\delta_0=200$ is very close to the total FoF mass of the halo ($M_{FoF} \simeq M_{\delta=200}$), which is particularly true for halos whose spherical overdensity mass is greater than 80\% of the FoF mass, the criterion used in the original selection of our halos (see section \ref{dark}).}

First of all, we observe the more massive the halos, the less spherical they are, $p(M)$ and $E(M)$ increase, while $b/a$ and $c/a$ decrease. Moreover, the curves obtained depend on the cosmology: for the halos considered, the median $p$ varies from 0.020 (low-mass halos) to 0.048 (high-mass halos) for the RPCDM model, and it ranges from 0.015 to 0.034 for the $\Lambda $CDM model and from 0.014 to 0.033 for the $w$CDM model. On average, the median $p$ for $\Lambda $CDM is about 25\% lower than that for RPCDM and 6\% higher than that for $w$CDM. A similar observation is made for triaxiality $T$ and $b/a$ curves. 

Secondly, the curves are ranked in ascending order of $\sigma_8$. The curves corresponding to the RPCDM model are far from the other curves, while the curves associated with the $w$CDM and $\Lambda$CDM models are very close, which corresponds to the differences in their respective $\sigma_8$ values (see \cref{param}).  When the $\sigma_8$ are small, the fluctuations at large redshift are small and halos tend to form more later. As halos tend moreover to sphericalize with time (due to the relaxation process, when the mass accretion process more or less stops) $\sigma_8$, halos formed later will therefore be less spherical. When $\sigma_8$ is low, the ellipticity, prolaticity, and triaxiality of the halos will therefore be higher, and the ratios of their axes $b/a$ and $c/a$ will be low. This is indeed what is observed in \cref{3,4}, the curves $E$, $p$ and $T$ corresponding to the RPCDM model are systematically above the curves corresponding to the $\Lambda$CDM, which are themselves above the curves corresponding to the $w$CDM.

Since the shape parameters of dark matter halos depend on the mass of the halos, but also on the cosmology according to the linear fluctuations of the cosmic matter field smoothed at 8 Mpc$/h$, it thus seems reasonable to express such parameters, more generally as a function of the fluctuations of the cosmic matter field smoothed on the scale of the mass of the halos. We explore this hypothesis in the next section.

\subsection{Shape of DM halos and linear fluctuations}
\label{shape}
\begin{figure}[h!]
    \includegraphics[width=8cm]{"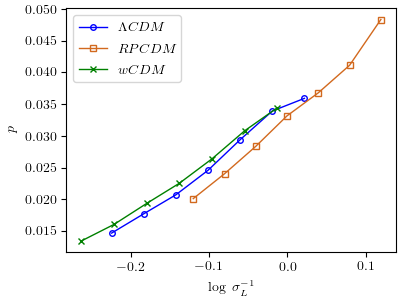"}
    \caption{Median prolaticity as a function of $\log \sigma_L(M)^{-1}$. The curves corresponding to the different dark energy models are closer to each other, comparatively to figure \ref{3} in which median prolaticity was plotted against halo mass. This proves that the linear power spectrum only partly explains the cosmological dependence of the relationship between halo mass and prolaticity. }
    \label{psl}
\end{figure}

\cite{Despali2014, Bonamigo2015,vega-ferrero2017} have studied the shape of halos at different redshifts, in the $\Lambda $CDM model. They expressed in this model, the halo shapes in terms of $\nu_L(M,z)=\frac{\delta_c(z=0)}{\sigma_L(M,z)}$, where $\sigma_L(M,z)$ is the amplitude of the mass-scale linear fluctuations of these halos at redshift $z$ and $\delta_c(z=0)$ is the critical spherical overdensity at $z=0$. In this work, the coefficients $\delta_c(z=0)$ of our three models are very close (see \cref{param}) to each other (we can use the usual fit $\delta_c(z=0)=1. 686\Omega_m^{0. 0055}$ to verify this). To test in the same way such universality, in terms of $\nu_L(M,z)$, at the same redshift ($z=0$) but this time for different cosmological models (Cf Table \ref{param}), it thus is sufficient to consider in our case the variable $\sigma_L$\footnote{We invite the reader to consult the results of \citet{alko2} where different models of modified gravity are studied with very different $\delta_c(z=0)$}. But, there are still two differences between our analysis and that of previous authors. Firstly, we have evaluated the shape properties of the DM halos detected by the FoF algorithm, whether relaxed or not, and secondly, the procedure used in \cite{Despali2012} to evaluate the shape of the halos in \cite{Despali2014,Bonamigo2015} differs from the one we use, as explained earlier in the section (\ref{triaxiality}). The results presented below can thus be considered as a robustness test, to the modifications mentioned above, of the proposal made by \cite{Despali2014} or \cite{Bonamigo2015}, of the universality of shape properties in terms of the linear fluctuations of the cosmic matter field,

From $M_{FoF}$ (the mass directly deduced from the FoF algorithm) of each halo, we calculated the linear fluctuation, $\sigma_L(M_{FoF})$. After arranging the halos in $\sigma_L$ bins and measuring the median prolaticity for the halos in each of these bins, we plot in \cref{psl} the median prolaticity of the halos as a function of $\sigma_L(M)$ for each cosmology. We observe that the curves approach but do not merge. The linear fluctuations $\sigma_L(M)$ are therefore not sufficient to "encode" all the cosmological dependence of the shape measurements in the case of our cosmological models where the dark energy model varies strongly through the coefficients $w$, $\sigma_8$ and $\Omega_m$, 

\subsection{Universality of the shape of Dark Matter and non-linear fluctuations}
\label{universality}
\begin{figure}[h!]
    \includegraphics[width=8cm]{"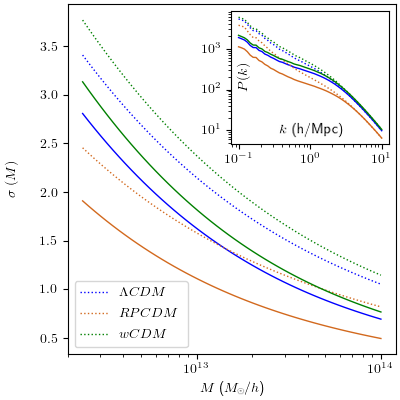"}
    \caption{The non-linear power spectra computed using all $z=0$ particles (dotted lines) or only those belonging to FoF halos, whose mass exceed $2.3\cdot 10^{12} M_{\odot}/h$ (solid lines). In the inner panel, the corresponding non-linear power spectra.   }
    \label{sih}
\end{figure}

\begin{figure}[h!]
    \includegraphics[width=8cm]{"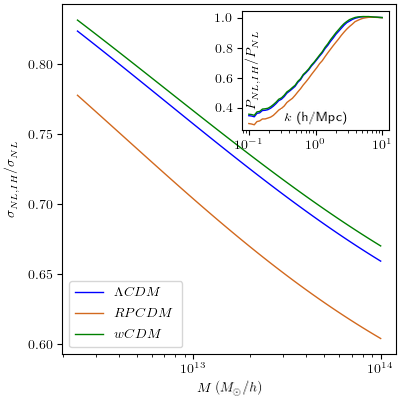"}
    \caption{Relative contribution of the matter inside the halos to the variance of the total non-linear power spectrum, $\sigma_{NL,IH}/\sigma_{NL}$. Manifestly, this ratio heavily depends on cosmology. It is worth noticing that even for low-mass halos, the matter inside the halos explains only 80 percent of the variance. In the inner panel, the corresponding ratio for the power spectra.  }
    \label{sihq}
\end{figure}

Re-expressed in terms of $\sigma_L(M)$, the cosmological dependence of the halo shape parameters has only been partially absorbed. However, since the shape properties were calculated not only on the particles inside the halos, which moreover evolved according to non-linear dynamics, it seems appropriate to re-express these properties not in terms of linear fluctuations of the total cosmic field but in terms of the non-linear fluctuations of the matter inside the halos, as introduced by \citet{Daalen2015, Pace2015}. We have therefore calculated the power spectrum of the density field smoothed over the mass scale of the halos (the smoothing chosen is again Gaussian smoothing, the density field was calculated from the particles inside the halos alone on a CIC grid of $3500^3$ cells) and we deduce the non-linear fluctuations $\sigma_{NL,IH}$. 

$\sigma_{NL}$ (the nonlinear fluctuations of the total matter field) and $\sigma_{NL,IH}$ (the nonlinear fluctuations of the cosmic matter inside halos), respectively $P_{NL}$ and $P_{NL,IH}$) are represented, as a function of masses $M_{FoF}$, respectively wavelengths $k$) on \cref{sih} and \cref{sihq}.

\begin{figure}[h!]
    \includegraphics[width=8cm]{"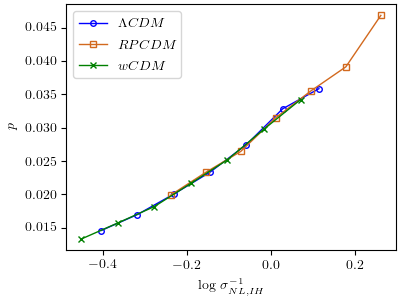"}
    \caption{Median prolaticity as a function of $\log \sigma_{NL,IH}(M)^{-1}$. The curves corresponding to the different dark energy models superpose, which means that all the cosmological information contained in (median) halo shape exactly corresponds to that carried by the non-linear power spectrum $P_{NL,IH}$}
    \label{psnl}
\end{figure}

$\sigma_{NL}$ dominates over $\sigma_{NL,IH}$ on mass scales corresponding to halo masses greater than $10^{13}$ $M_{\odot}$, and the fluctuations associated with cosmic matter inside the halos account for only eighty percent of the fluctuations of total matter field. This ratio varies with mass and the cosmological dependence does not reduce to a constant factor of proportionality. We then plot on \cref{psnl} the median halo prolaticity as a function of $\sigma_{NL,IH}$ for the $\Lambda$CDM, RPCDM and $w$CDM cosmological models. The curves corresponding to the different cosmological models are then fully superimposed very precisely, less than $3\%$ of difference. The relationship between the median $p$ of the halos and $\sigma_{NL,IH}^{-1}$ become thus universal, it no longer depends on the cosmological model. $\sigma_{NL,IH}$ or, in other words, the non-linear power spectrum of matter inside the halos contains all the cosmological dependence of the shape properties of the halos.

\begin{figure*}
    \subfloat[Median middle-to-major axis ratio versus $\log \sigma_{NL,IH}(M)^{-1}$]{\includegraphics[width=8cm]{"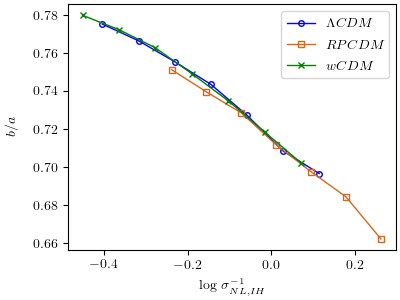"}}
    \subfloat[Median triaxiality versus $\log \sigma_{NL,IH}(M)^{-1}$]{\includegraphics[width=8cm]{"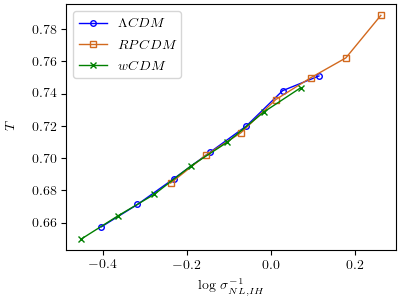"}}\\
    \subfloat[ Median minor-to-major axis ratio versus $\log \sigma_{NL,IH}(M)^{-1}$]{\includegraphics[width=8cm]{"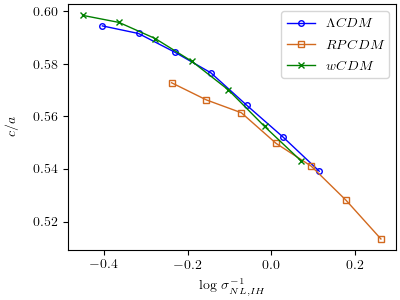"}}
    \subfloat[ Median ellipticity versus $\log \sigma_{NL,IH}(M)^{-1}$]
    {\includegraphics[width=8cm]{"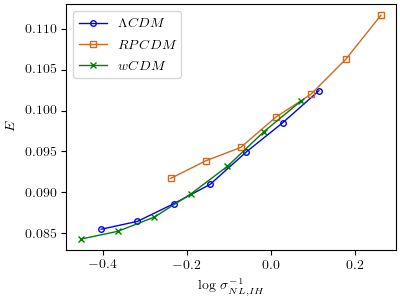"}}
    \caption{Dependence of other halos shape parameters on cosmology and non-linear fluctuations heights. }
    \label{autrsnlih}
\end{figure*}

\cref{autrsnlih} shows that this conclusion is also very well verified for $b/a$ and $T$. We also plot the curves for $E$ and $c/a$. This time, a slight deviation appears at low masses, particularly for the RPCDM model. We do not consider this to be significant, as the low number of particles along the third axis of the inertia tensor ($c$), precisely for low-mass halos, biases the measurement of these shape parameters and artificially favors a higher ellipticity. We discuss in \cref{appb}, the influence of the number of particles per DM halo on the accuracy of the measure of shape parameters. 

But does such universality in terms of $\sigma_{NL-IH}$ persist in terms of non-linear fluctuations calculated now over the entire cosmic field? We now discuss this point. 

\begin{figure}[h!]
    \includegraphics[width=8cm]{"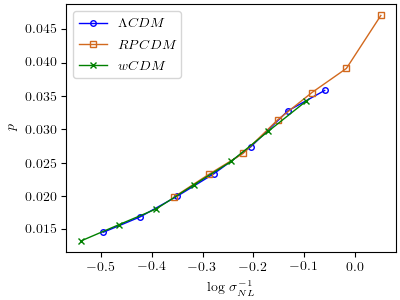"}
    \caption{Median prolaticity as a function of $\log \sigma_{NL}^{-1}$. The curves corresponding to the different cosmological models superpose, as they did when binning in $\log \sigma_{NL,IH}^{-1}$. This suggests the existence of a cosmological independent mapping between variances of total and in-halos non-linear power spectra. }
    \label{pIH}
\end{figure}
\begin{figure*}[h!]
    \subfloat[][Median middle-to-major axis ratio versus $\log \sigma_{NL}(M)^{-1}$]{\includegraphics[width=8cm]{"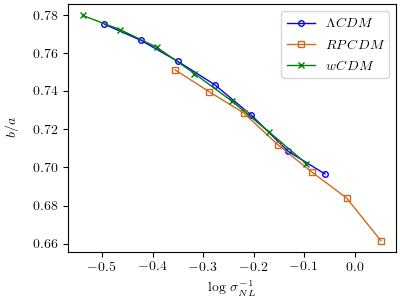"}}
    \subfloat[][Median triaxiality versus $\log\sigma_{NL}(M)^{-1}$]{\includegraphics[width=8cm]{"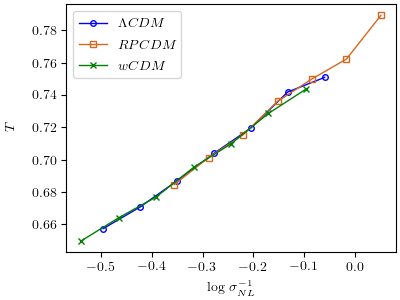"}}\hfill{"}
    \subfloat[][Median minor-to-major axis ratio versus $\log \sigma_{NL}(M)^{-1}$]{\includegraphics[width=8cm]{"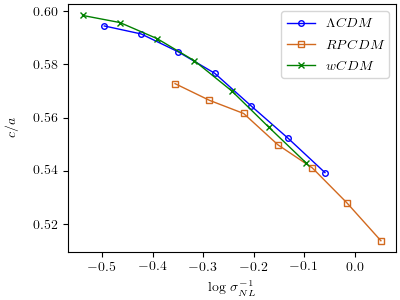"}}
    \subfloat[][ Median ellipticity versus $\log \sigma_{NL}(M)^{-1}$]{\includegraphics[width=8cm]{"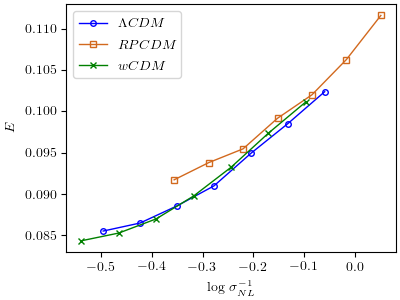"}}
    \caption{Dependence of other dark matter halo shape parameters on the non-linear fluctuations of the cosmic matter field inside the halos for the three cosmological models.}
    \label{autrsnl}
\end{figure*}

Figure \cref{pIH} now shows the prolaticity of the halos in terms of the non-linear fluctuations $\sigma_{NL}$ calculated on the whole cosmic matter field for the $\Lambda$CDM, RPCDM, and $w$CDM cosmological models. The different curves are again fully superimposed.  The results are similar for the other shape parameters: \cref{autrsnl} shows that their median values also do not vary with cosmology in terms of $\sigma_{NL}$. 

\begin{figure}[h!]
    \includegraphics[width=8cm]{"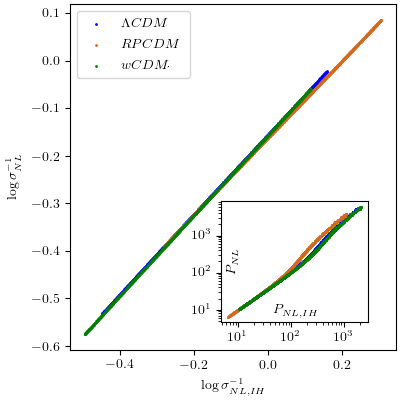"}
    \caption{Mapping between variances of total and in-halos non-linear power spectra. For the mass scales probed by the simulation, $M\in [2.3\cdot 10^{12}, 10^{14}]  M_{\odot}/h$, it consists in the scatter plots $\{(\log \sigma_{NL,IH}^{-1}(M),\log \sigma_{NL}^{-1}(M))\}$ for each cosmology. Those superpose, meaning that $\sigma_{NL}$ is a single-valued universal function of $\sigma_{NL,IH}$. This function is manifestly a power law.   }
    \label{fif}
\end{figure}

Since the prolaticity $p[\sigma_{NL,IH}]$ and $p[\sigma_{NL}]$ in terms of $\sigma_{NL,IH}$ and $\sigma_{NL}$ respectively are both independent of cosmology, there is necessarily a cosmology-independent correspondence between the non-linear fluctuations of matter in halos, $\sigma_{NL,IH}$ and the non-linear fluctuations of all cosmic matter $\sigma_{NL}$. 

We plot in \cref{fif} in the mass domain $M\in [2\cdot 10^{12},10^{14}]$ the non-linear fluctuations of the total cosmic matter field as a function of the non-linear fluctuations of the interior matter field of the halos. We again clearly observe that the three curves coincide, which confirms the existence of a cosmology-independent relationship between $\sigma_{NL,IH}$ and $\sigma_{NL}$. We also note that the correspondence is very well approximated by a power law.

\begin{figure}[h!]
    \includegraphics[width=8cm]{"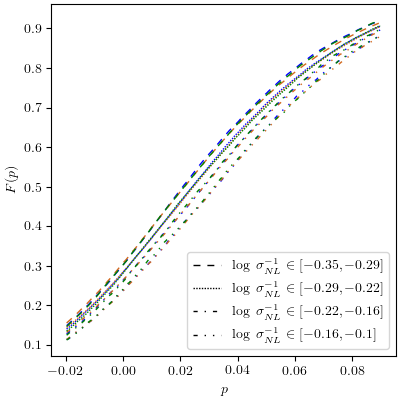"}
    \caption{Dependence of prolaticity distribution on cosmology and $\log \sigma_{NL}^{-1}$. We define four bins of $\log \sigma_{NL}^{-1}$ and then compute the cumulative distribution function of prolaticity, in each bin. The curves correspond to three cosmological models superposes, not only for $F=0.5$ (median prolaticity) but all along the distribution (between the first and ninth decile). Hence, one can say that the whole prolaticity distribution is related to the non-linear fluctuations in a non-cosmology-dependent way. }
    \label{cdf}
\end{figure}
    
\begin{figure}[h!]
    \includegraphics[width=8cm]{"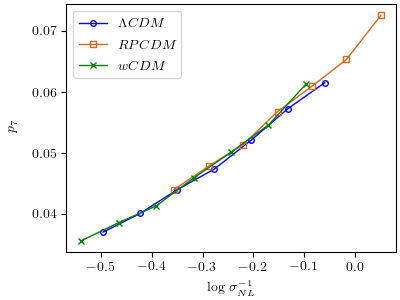"}
    \caption{Seventh decile prolaticity as a function of $\log \sigma_{NL}(M)^{-1}$. The curves corresponding to the different dark energy models superpose, meaning the universality is not only verified for median prolaticity but for the whole of prolaticity distribution. }
    \label{p70}
\end{figure}
A similar comparison is presented in the internal panel of \cref{fif} for $P_{NL}$ and $P_{NL,IH}$.  Here again, there is a universal relationship between these two quantities as long as $P_{NL,IH}<10^{2}$, which corresponds to $k\geq 2 \,h\textrm{Mpc}$. The lower modes correspond to a distance greater than 6 \textrm{Mpc}/h, they describe correlations between the DM halos. On these last scales, a cosmological dependence appears.

\begin{figure*}[h!]
    \includegraphics[width=17cm]{"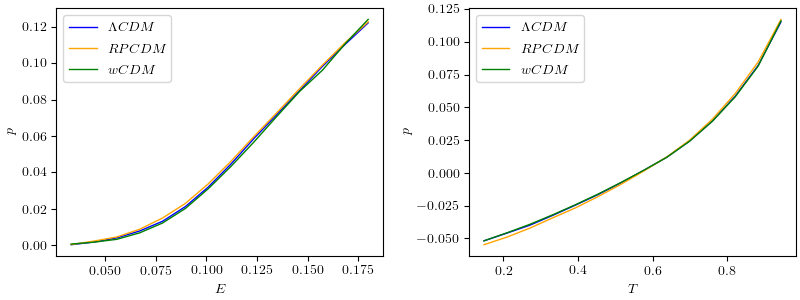"}
    \caption{Median prolaticity as a function of ellipticity (left) or triaxiality (right). The relationships between the various shape parameters do not depend on the cosmological model. It is a direct consequence of the fact that there exists a variable ($\log \sigma_{NL}^{-1}$) to which median shape parameters are universally related (see main text). }
    \label{eTp}
\end{figure*}
Up to now, the universal and cosmology-independent relation between the shape properties of DM halos and the non-linear fluctuations of the cosmic matter field has been highlighted for the \textbf{median} quantities calculated on the set halos with a given $\sigma_{NL}$. But what about the rest of the halo prolaticity distribution? Can we say that the non-linear fluctuations are sufficient to determine the entire prolaticity distribution of DM halos? To answer this question, we calculated the cumulative distribution function (CDF) of the halo prolaticity for given $\sigma_{NL}$ slices. We chose sufficiently large $\sigma_{NL}$ intervals to be sure of having enough halos and therefore sufficient statistics. \cref{cdf} shows that, in each bin, the CDF of all cosmological models coincides again. In terms of the rms of the non-linear power spectrum of the cosmic matter field, the whole cumulative distribution function of $p$ is therefore well independent of cosmology. We also plot, for example, the seventh decile of the prolaticity distribution as a function of $\sigma_{NL}$. The curves of this seventh decile are again superimposed for all cosmologies, as we can see in \cref{p70}. 

\subsection{Universality of shape parameters relations} 
\label{relations}
Finally, \cite{Despali2014} also notes that there is a relationship, independent of $z$, between the median ellipticity $E$ and the median prolaticity $p$ of the DM halos they have studied. We obtain a similar result but this time at redshift $z=0$ fixed, for different cosmological models. Indeed, \cref{eTp} presents for the population of halos of our three cosmological models at $z=0$, the median $p$ in each bin of $E$ (resp. $T$), the relations again obtained do not depend on the cosmological model. However, the results of the previous section now easily explain such a result. The median prolaticity and median ellipticity (respectively the triaxiality $T$) of our halo populations are, expressed in terms of $\sigma_{NL}$, invariant functions according to the cosmology, and these functions are strictly monotonic and therefore bijective, necessarily implies that there is a universal relation between prolaticity and ellipticity (respectively triaxiality) as observed on \ref{eTp}. Formally, this can be rewritten as follows: if $f$ and $g$ are the median functions of prolaticity and ellipticity (respectively triaxiality) in terms of $\sigma_{NL}$, then the universal relation of prolaticity as a function of ellipticity (respectively triaxiality) is effectively universal $p(E)=g(f^{-1}(E))$ (respectively $p(T)=g(f^{-1}(T))$).

\section{Reconstruction of non-linear fluctuations from properties of DM halos shape}
\label{reconstruction}
\subsection{Reconstruction of rms non-linear fluctuations }
\label{reconstructionrms}
We have shown that it is sufficient to re-express the distribution of dark matter halo shape parameters in terms of non-linear $\sigma_{NL}$ fluctuations in the cosmic matter field to remove any cosmological dependence. Based on this invariance and the dependence of the distribution of dark matter halo shape parameters on their mass, which can otherwise be measured, we will now reconstruct the non-linear fluctuations of the cosmic matter field in which these dark matter halos formed. 

From the masses and prolaticity (or any other shape parameter) of a set of dark matter halos, we deduce the median prolaticity in mass bins of these halos as it has been done in figures \ref{3} and \ref{4}. As the relationship between median prolaticity and $\sigma_{NL}$ is universal, i.e. cosmologically invariant, we deduce $\sigma_{NL}(M)$ from a simple inversion of the latter relationship, after re-expressing prolaticity in terms of mass as measured previously. In other words, from the universal relation $p(\sigma_{NL})$ we deduce $\sigma_{NL}(p)$, we then re-express this relation in terms of mass, using the specific relation of each cosmological model, which was measured previously $p(M)$ and we thus deduce $\sigma_{NL}(p(M)) \equiv \sigma_{NL}(M)$.

\begin{figure}[h!]
    \includegraphics[width=8cm]{"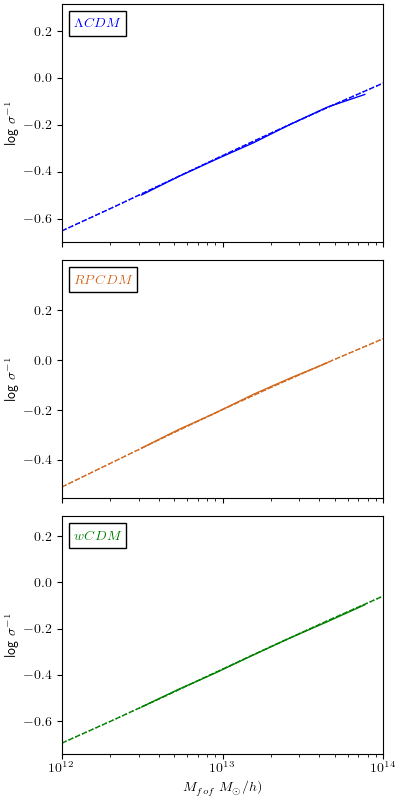"}
     \caption{Reconstruction of non-linear fluctuations in the rms of the cosmic matter field. The solid curves represent $\log\sigma_{NL}^{-1}$ as a function of $M$ and are reconstructed from measurements of median prolaticity deduced from the shape of the halos (as a function of the halo mass) and from the universal relations between median prolaticity and $\log\sigma_{NL}^{-1}$. The agreement with the expected non-linear fluctuations (dotted lines) of the cosmic matter field derived from simulations and calculated with Powergrid at $z=0$ is excellent. We have also plotted (dashed lines) $\log\sigma_{L}^{-1}(M)$ for comparison: the shape measurements undoubtedly give information about the non-linear fluctuations of the cosmic matter champ that are distinct from the linear fluctuations on the scale of the dark matter halo.}
     \label{snl}
    \end{figure}

We have applied this procedure to our sets of DM halos. In practice, the universal curve $p(\sigma_{NL})$ is chosen by averaging the three almost indistinguishable \cref{psnl} curves. From measuring prolaticity as a function of mass for all halos of a particular cosmological model, we deduce $\sigma_{NL}(M)$ for that model. We repeat this procedure for our three cosmological models, and the results are given in \cref{snl}. The dotted lines correspond to the expected $\sigma_{NL}(M)$ deduced from the non-linear power spectrum computed on the CIC density field at $z=0$. The correspondence between the expected mass dependence and the mass dependence of $\sigma_{NL}$ deduced from measurements of the shape of the halos (contiguous lines) is very satisfactory.

\subsection{Reconstruction of non-linear power spectrum}
\label{reconstructionspectrum}
We now deduce the non-linear power spectrum of the cosmic matter field from the previous reconstruction of $\sigma_{NL}(M)$. Assuming a Gaussian window as the filter for the cosmic matter field, $\sigma_{NL}(M)$ is then defined as follows 
\begin{equation}
    2\pi^2{\sigma_{NL}^2}(R)=\int_0^{+\infty}k^2P_{NL}(k)\exp\left(-\frac{R^2k^2}{5}\right)\mathrm{d}k
\end{equation}
We now apply the change of variable  
\begin{equation*}
    f(x)=2\pi^2{\sigma_{NL}^2}(\sqrt{5x}),\;\;\;y=k^2\,\,\,\textrm{and}\,\,\,g(y)=\frac{\sqrt{y}}{2}{P}_{NL}(\sqrt{y})
\end{equation*} 
$f$ is then the Laplace transform of $g$
\begin{equation}
f(x)=\int_0^{+\infty}g(y)\exp(-xy)\mathrm{d}y \equiv \mathcal{L}[g](x)\label{fg}
\end{equation}
By supposing $\sigma_{NL}(M)$ as a power law in terms of $M$, i.e. $\sigma_{NL}(M)=10^{\alpha}M^{\beta}$ (which is a very good approximation as it can be observed in figure \ref{rms-pk} and \ref{snl}), we obtain 
\begin{equation}
   f(x)=2\pi^2\cdot 10^{2v}5^{u} x^{u }
\end{equation} 
where $u=3\beta<0$, $v=\alpha+\beta\log_{10}(\mathrm{vol})$ and $\mathrm{vol}=\frac{4}{3}\pi \rho_c\Omega_m$.

The parameters $\alpha$ and $\beta$ are estimated from the $\sigma_{NL}$ that we had reconstructed solely from the data, as presented in section \ref{reconstructionrms} (figure \ref{snl}). These parameters are listed in table \ref{tab:alphabeta}. 
\begin{table}[h!]
    \centering
    \begin{tabular}{|c||c|c|c|c|}
    \hline
    Cosmology & $\alpha$ & $\beta$ &$n$ & $b_n$ \\
    \hline\hline 
    $\Lambda$CDM & $4.39818$ & $-0.31243$&$-1.12541$ & $361.332$ \\
    \hline
    RPCDM & $4.05868$ & $-0.29685$&$-1.21889$ & $179.451$ \\
    \hline
    $w$CDM & $4.50881$ & $-0.31787$&$-1.09276$ & $444.412$
    \\
    \hline
  \end{tabular}
    \caption{$\alpha$ and $\beta$ power law parameters (see text) modeling the  rms nonlinear fluctuations $\sigma_{NL}$ of the figure \ref{snl}. The parameters $n$ and $b_n$ are calculated from equation \ref{bnn} by supposing a common value $\Omega_m = 0.25$ for all cosmological models}
    \label{tab:alphabeta}
\end{table}
We deduce 
\begin{eqnarray*}
g(y)&=&\mathcal{L}^{-1}[f](y)\\&=&2\pi^2\cdot 10^{2v}5^{u} \mathcal{L}^{-1}[{x\mapsto x^{u}}](y)\\&=&2\pi^2 \cdot 10^{2v}5^{u} \cdot  \frac{y^{-1-u}}{\Gamma(-u)} 
\end{eqnarray*}
and finally, we get,  
\begin{equation}
    {P}_{NL}(\sqrt{y})=4\pi^2 10^{2v}5^{u} \frac{y^{-u-3/2}}{\Gamma(-u)}
\end{equation}
or 
\begin{equation}
    {P_{NL}(k)=\frac{\left(2\pi\cdot 10^{v}\right)^2}{5^{-u}\Gamma(-u)} k^{-2u-3}\equiv b_n k^n }
\end{equation}
with 
\begin{eqnarray}
\label{bnn}
  n=-6 \beta -3
  &\mathrm{and}&
  b_n = \frac{4^{\alpha+1}\pi^2 5^{2\alpha+3\beta}\left({\frac{4}{3}\pi\rho_c\Omega_m}\right)^{2\beta}}{\Gamma(-3\beta)} 
\end{eqnarray}
$P_{NL}$ is obviously a power law in terms of modes $k$. The power law exponent, $n$, is unchanged whatever the window shape as the filter for the cosmic matter field. $n$ and $b_n$ are specific to each cosmological model. $n$ is deduced only from data, $b_n$ is  deduced from data and from some prior on $\Omega_m$. $n$ and $b_n$ are presented in table \ref{tab:alphabeta}.

\begin{figure*}[h!]
    \includegraphics[width=17cm]{"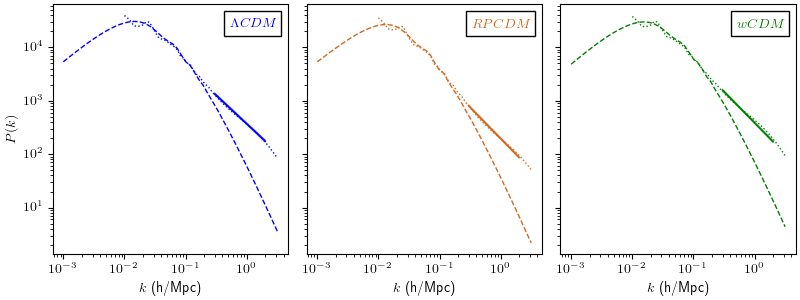"}
    \caption{Reconstruction of non-linear power spectrum. For each model, we reconstruct $P_{NL}(k)$ (solid lines) from the estimation of $\sigma_{NL}(M)$ deduced from shape measures of DM halos at $z=0$. The thickness of the solid line reflects an imprecision of $5\%$ on the true value of the density parameter $\Omega_m$ assumed from independent measurements. The expected non-linear power spectrum and linear power spectrum deduced from the cosmic matter field of numerical simulation are respectively in dotted line, and dashed line. The agreement between the non-linear power spectrum and the reconstructed power spectrum is excellent.}
    \label{pk}
\end{figure*}

The comparison between the reconstructed power spectrum of the cosmic matter field deduced from the shape of DM halos (solid lines) and the power spectrum deduced directly from the cosmic matter field of numerical simulations (dotted lines) are presented in \cref{pk} for the three cosmological models. The thickness of the solid line reflects an imprecision of $5\%$ on the true value of the density parameter $\Omega_m$ assumed from independent measurements. 

The power spectrum on the scale corresponding to DM halos is precisely reconstructed for the three cosmological models.
 
\subsection{Estimation of $\sigma_8$ cosmological parameter }
\label{estimation}

\begin{figure}[h!]
    \includegraphics[width=8cm]{"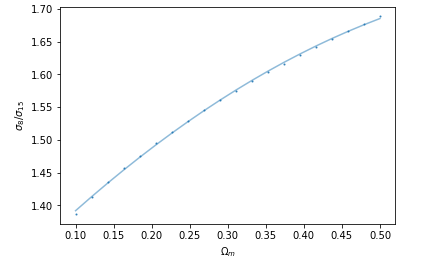"}
    \caption{$\sigma_L(R=8\textrm{Mpc}/h)/\sigma_L(R=15\textrm{Mpc}/h)$ for various $\Omega_m$, from CAMB. Only $\Omega_m$ was varied, the other cosmological parameters $n_s,\Omega_b...$ are taken identical to WMAP7 constraints, see \cref{param}.  The result is almost insensitive to DE parameters $(w_0,w_a)$}
    \label{s8}
\end{figure}

We now estimate $\sigma_8$. First, we remark (\cref{rms-pk}) that $\sigma_L(M)={\sigma_{NL}}(M)$ for $M>10^{15}$ solar mass. Second, from the linear power spectra obtained by using CAMB software \citep{camb} for cosmological models, where we considered $w_0\in [-1.5,-01.75]$, $w_a\in[0,0.5]$, $1-\Omega_\Lambda \in [0.2,0.3]$ and, the parameters that are common to our three models unchanged,  $\Omega_bh^2=0.02258$, $n_s=0.96$ and $H_0=72$ km/s/\textrm{Mpc}, we deduce  
\begin{equation}
    \label{fits8s15}
    \sigma_L(M_8)=\sigma_L(M_{15})(-0.74086\Omega_m^2+1.17774\Omega_m+1.28147)
\end{equation}
where $M_8=\frac{4\pi 8^3\Omega_m\rho_c}{3}$. Such an adjustment (figure \ref{s8}) is a function of $\Omega_m$ because the ratio $\sigma_L(M_8)/\sigma_L(M_{15})$ is a ratio of two integrals of the transfer function, which itself depends on $\Omega_m$. Around $w\sim -1$ the dependence of the transfer functions is very low \citep{Ma99}. Such an adjustment holds to less than one percent. 

\begin{figure}[h!]
    \includegraphics[width=8cm]{"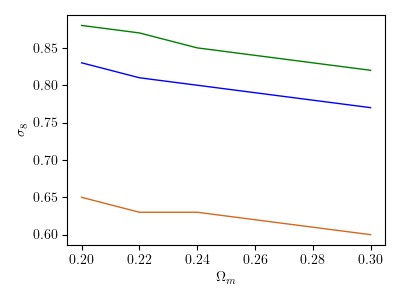"}
    \caption{Estimation of the linear cosmological parameter $\sigma_8$ from prolaticity measures, assuming different values of $\Omega_m$ in the range $[0.2,0.3]$. }
    \label{oms8}
\end{figure}

If one admits moreover, the extension to $M=M_{15}$ of the relationship ${\sigma_{NL}}(M)=10^\alpha M^\beta$ i.e. \begin{equation}
{\sigma_{NL}}(M_{15})\approx 10^\alpha \left(\frac{4}{3}\pi 15^3\rho_c \Omega_m\right)^\beta
\end{equation}
we get 
\begin{eqnarray*}
    \sigma_8[\Omega_m]=10^{\alpha} \left(\frac{4}{3}\pi 15^3\rho_c \Omega_m\right)^{\beta}(-0.74086\Omega_m^2&+&1.17774\Omega_m\\&+&1.28147)   
\end{eqnarray*}

where $\sigma_8[\Omega_m]$ is thus very weakly dependent on $\Omega_m$ varying in the range (0.2-0.3) because $\beta \simeq -0.3$ and consequently we can affirm that halo shapes are then a probe of $\sigma_8$ but cannot be used for constraining $\Omega_m$. We plot in figure \ref{oms8} the function $\sigma_8[\Omega_m]$ estimated through median $p[M]$ and universal $p[{\sigma_{NL}}]$ linear regressions. 

By supposing a homogeneous prior of $\Omega_m$ over the range $[0.2,0.3]$, one gets an estimated $\overline{\sigma_8}=10\int_{0.2}^{0.3}\sigma_8[\Omega_m]\mathrm{d}\Omega_m$. Compared to the “exact” $\sigma_8$ (computed through direct linear power spectrum measure), the result is extremely satisfying, see Table \ref{s8s8}. 

\begin{table}[h!]
    \begin{tabular}{|c||c|c|c|}
    \hline
    Model & $\Lambda $CDM & RPCDM  & $w $CDM \\ 
    \hline
    \hline
    Exact $\sigma_8^{Gauss}$  &0.83 &0.68 & 0.88 \\
    \hline
    Estimated $\overline{\sigma_8^{Gauss}}$&0.79 &0.62 & 0.85\\
    \hline
\end{tabular}
    \caption{Estimation of $\sigma_8$ parameter from mass and shape measures in dark energy simulations (see main text). The exact value, taken from direct integration of the linear power spectrum is also reported for comparison.}
    \label{s8s8}
\end{table}

\section{Conclusions}
\label{conclusion}
In this article, we present a study concerning the fundamental relationship between the morphology of DM halos and the cosmological models in which they formed, using data from the dark energy universe simulations of three different realistic dark energy models. The cosmological models of the DEUS simulations are a flat $\Lambda $CDM model, a quintessence model with Ratra-Peebles potential (RPCDM) or equivalently to a constant equation of state $w=-0.8$, and a phantom dark energy model with $w=-1.2$ ($w$CDM). 

We first present a method for evaluating the shape of dark matter halos detected in numerical simulations, which corrects for the effects due to the presence of numerical resolution-dependent substructures and adopts an isodensity measure to characterize the shape of dark matter halos, thus extending its applicability to other methods of detecting dark matter halos than the FoF algorithm. As long as the edge of the detected halos does not intersect the isodensity under consideration (in practice $\delta=200$), any algorithm is suitable. Using such a robust method, we have shown that the shape of the halos is strongly influenced by the underlying cosmology. Halos tend to be more oblate and spherical in high $\sigma_8$ models, i.e. for a more structured cosmic matter field with more collapsed halos.

While it is true that the linear power spectrum, through $\sigma_8$, basically controls the asphericity of the halos, we have shown that it is possible to completely reduce the cosmological dependence of the distribution of the shape of the halos by taking into account the non-linear dynamics that led to their formation. Indeed, the distributions of the prolaticity, triaxiality, and ellipticity of dark matter halos at a given level of non-linear fluctuations in the cosmic matter field are perfectly independent of cosmology. Thus, while these shape parameters are cosmology-dependent when considered as a function of mass, they are cosmology-independent when considered as a function of the variance of the power spectrum of the cosmic matter field, smoothed at the scale of the halos. This reflects the central role of non-linear dynamics in the evolution of the sphericity of dark matter halos. In fact, the higher the variance of non-linear fluctuations in the cosmic matter field, the more structured the Universe and the more collapsed the halos that have formed. Since Newtonian iso-potentials are systematically more spherical than the iso-densities of the tri-axial halos that form, the halos become more spherical as they collapse. The resulting cosmological invariance relation is not subject to the selection of relaxed halos. We have also shown that this universality result persists if we substitute the variance of the non-linear fluctuations of the cosmic matter field contained only in the halos for the variance of the non-linear fluctuations of the total cosmic field. This then leads to a correspondence between these two measures of variance and the two associated power spectra, again independent of cosmology, which disappears, however, at the scales corresponding to the distribution of the DM halos between them, which again becomes dependent on cosmology.

We then show that the distribution of the shape parameters of the DM halos according to their mass is sufficient to reconstruct the non-linear power spectrum of the cosmic matter field in which these halos formed. The excellent agreement between the reconstructed power spectrum and the non-linear power spectrum measured directly in the numerical simulations demonstrates the robustness of the invariance on the shape of the halos that have been highlighted. We also find, with a good approximation, the value of the cosmological parameter $\sigma_8$ of the cosmological model in which the cosmic matter field evolved and the halos formed. 

The results presented in this article have therefore highlighted the importance of analyzing the shape of dark matter halos as a powerful tool for probing the non-linear dynamics of the cosmic matter field. They have shown the central role of non-linear dynamics in the increase in the sphericity of DM halos during their formation. They also open up prospects for future developments.

Does this cosmological invariance persist within the framework of modified gravity models, alternatives to Newton's and Einstein's theory of gravity? A generalization or, on the contrary, a break in this invariance could reveal new aspects of it, enriching our understanding of the fundamental laws that govern the universe. We will examine this question in a forthcoming paper (Koskas \& Alimi 2024). On the other hand, a more original nature of dark matter, which would imply a process of anisotropic formation of cosmic structures on the scale of DM halos, could also constitute fertile ground for testing the robustness and scope of the cosmological invariance demonstrated for the $w$CDM models tested in this article. Can the universality of the invariance, or on the contrary its breakage in fuzzy dark matter models, shed new light on the intrinsic nature of dark matter and the mechanisms of formation of cosmic structures in this context? In addition, it would also be essential to consider the influence of the presence of baryons, more particularly in lower-mass galactic structures where the complex interaction between dark matter and baryonic matter and its impact on the morphology of cosmic structures could reveal unexplored aspects of a possible cosmological invariance.  

Recent works \citep{Schneider_2019} tend to show that the power spectrum is sensitive to baryonic feedback (for certain feedback models only), including for scales $k\sim 0.1 Mpc/h$. But as long as the presence of baryons does not strongly modify the isotropization of the collapse of cosmic matter, it is possible that this presence should not modify the universality result established in our work.  Either, as shown by other work \citep{2022MNRAS.515.2681C}, because the baryonic feedback modifies the shape of the halos in a way equivalent to the impact that these baryons induce on the power spectrum of cosmic matter, or because the presence of the baryons does not sufficiently modify the isotropization of the collapse and therefore leaves the power spectrum and the shape of the halos almost unchanged.  The relationship between shape and power spectrum then remains independent of (non-baryonic) cosmological parameters, but the question for halos with masses less than $10^{12} M_\odot$ remains effectively open.  

To sum up, this article not only charts a new course in our understanding of dark matter and its cosmic dynamics, but it also lays the foundations for a profoundly rewarding multidisciplinary exploration of the universe. The prospects offered by the extension of our study to alternative cosmological models and the dark matter-baryon interaction promise exciting advances in our quest to understand the universe. It is an invitation to push back the frontiers of cosmology, embrace complexity, and boldly pursue our exploration of the mysteries of cosmic space.

In terms of prospects, it is also crucial to develop robust methodologies for testing this cosmological invariance on observational data. The difficulty lies in accurately measuring the shape of galaxy clusters and other cosmic structures, as well as their DM halos. In this respect, weak gravitational lensing techniques probably offer a promising way forward. By analyzing the distortions undergone by the light from background galaxies, weak gravitational lensing can reveal the mass distribution of foreground structures such as the DM halos of galaxy clusters. This method therefore provides an indirect but powerful means of mapping the distribution of the cosmic matter field and thus testing the cosmological invariance highlighted in our study. In addition, maps of the gas in galaxy clusters, obtained by X-ray observations or by the Sunyaev-Zel'dovich effect, may represent another promising approach. These methods could also provide complementary information on the distribution of matter in the universe, offering a unique perspective for understanding the shape and distribution of DM halos and retesting a possible fundamental cosmological invariance. 

Finally, the study presented in this article opens up prospects for observational and physical cosmology. It calls for an interdisciplinary theoretical and observational study to overcome the technical challenges associated with measuring the shape of halos and to fully exploit the potential of dark matter halo shape analysis as a powerful tool for probing some of the mysteries of the universe through the fundamental cosmological invariance presented in this paper.

\section*{Acknowledgments}
R.K. is dedicated to the loving memory of William Nessim Smadja (CNRS) who passed away during the writing of this article. \\
The authors thank Carlo Schimd (LAM), Emmanuel Nezri (LAM), and Yann Rasera (LUTH) for fruitful discussions. \\
This work was granted access to HPC resources of TGCC, CCRT, and IDRIS through allocations made by GENCI (Grand \'Equipement National de Calcul Intensif) in the framework
of the “Grand Challenges” DEUS and DEUSS. \\
All the calculations of this paper are based on C++ and Python codes written by the authors, using Numpy \citep{numpy}, Matplotlib \citep{matplotlib}, Scipy \citep{scipy}, and Pynverse\footnote{\url{https://github.com/alvarosg/pynverse/tree/master}} libraries, and were executed with GNU-Parallel \citep{parallel}

\bibliographystyle{aa.bst}
\bibliography{alimiandkoskas.bib}

\appendix
\section{Two-dimensional projections and Redshift evolution}
\begin{figure*}[h!]
    \subfloat{\includegraphics[width=8.5cm]{"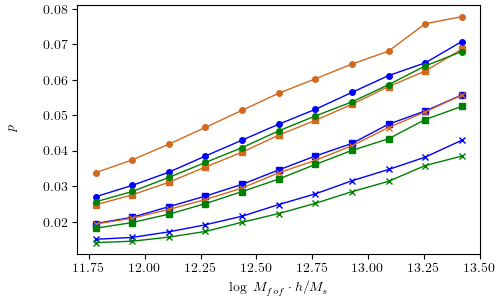"}}
    \subfloat{\includegraphics[width=8.5cm]{"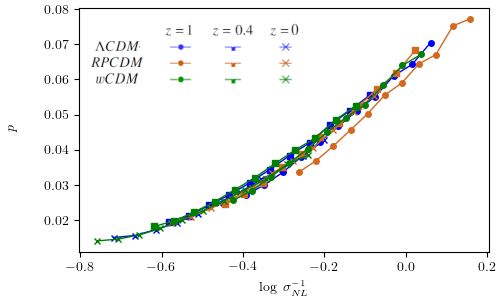"}}
     \caption{Median prolaticity for all cosmologies and various redshifts, as a function of halo mass (left) and non-linear variance (right). The variable change $M\rightarrow \log\sigma_{NL}^{-1}$ compensates for the cosmological and redshift dependence, and all the curves coincide. Universality according to the cosmology \textit{implies} universality in redshift. }
     \label{tz}
\end{figure*}

The cosmological invariance of the shape properties of the halos as a function of non-linear fluctuations in the cosmic field is also found at all redshifts. Indeed, from halos in the three cosmological models $w$CDM, $\Lambda $CDM and RPCDM but this time at redshifts $z=1,0.4,0$ (see \cref{tabzz}), we have calculated the median prolaticity according to the mass of the halos and according to $\sigma_{NL}(M,z)$. The prolaticities were calculated by considering all the particles belonging to the halos as identified by FoF. Indeed, the main purpose of the substructure removal procedure (and the resulting selection criteria, section \ref{triaxiality}) was to remove the effects of resolution on the shape of the halos. If we reason with fixed resolution, these steps are no longer strictly required, and the shape of the FoF halos without treatment is still a function of $\sigma_{NL}$ independent of the cosmology (although necessarily dependent on the resolution).  \\
The result is presented in \cref{tz}. 

Once again, cosmological invariance is found. $\sigma_{NL}$ calculated at the adapted redshift contains all the cosmological dependence of the shape properties of the DM halos.

\begin{table*}[h!]
\begin{tabular}{|c|c|c|c|}
\hline
Cosmology   &   Redshift    &   halos $\geq 100$ particles    &   halos $\geq 1000$ particles\\
\hline
\hline
$\Lambda$CDM& $z=1.0$ & 2,888,709 & 309,507\\
$\Lambda$CDM& $z=0.4$ & 3,056,944 & 399,708\\
$\Lambda$CDM& $z=0.0$ & 3,045,305 & 441,934\\
\hline
RPCDM       & $z=1.0$ & 2,518,366 & 223,845\\
RPCDM       & $z=0.4$ & 2,926,349 & 338,873\\
RPCDM       & $z=0.0$ & 3,066,884 & 405,050\\
\hline
$w$CDM      & $z=1.0$ & 2,980,683 & 329,397\\
$w$CDM      & $z=0.4$ & 3,075,878 & 413,170\\
$w$CDM      & $z=0.0$ & 3,015,407 & 446,818\\
\hline
\end{tabular}
\caption{Number of halos at the various redshifts}
\label{tabzz}
\end{table*}

\begin{figure*}[h!]
        \includegraphics[width=17cm]{"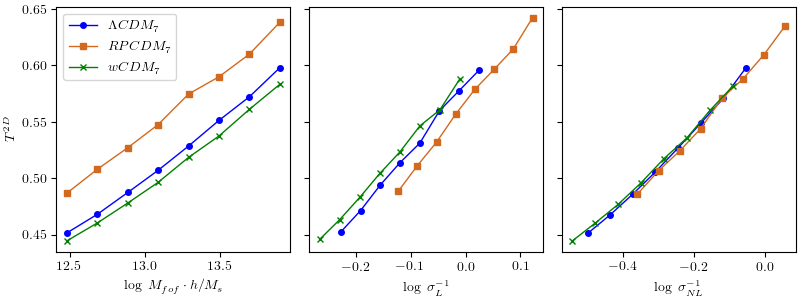"}
    \caption{Median squared eccentricity as a function of mass (left), linear variance (middle), and non-linear variance (right). After projecting all the halos onto the simulation  $(x,y)$ plane, we diagonalize the two-dimensional mass tensor of each projected halo. The median eccentricity is still invariant according to cosmology when expressed in $\log\sigma_{NL}^{-1}$ bins.  }        \label{tFoF2d}
\end{figure*}

The cosmological invariance of the shape properties of DM halos as a function of non-linear fluctuations in the cosmic matter field is preserved in the two-dimensional projected halos. Indeed, from the FoF halos of the numerical simulations we have only retained the projection of the particles making up these halos in the $(x,y)$ plane of the simulation (the choice of the plane is not important and the result obtained is preserved whatever the plane). From these projected positions, we then calculate the associated two-dimensional inertia tensor. Let $a^2\geq b^2$ be its eigenvalues, the square of the eccentricity is then given by $E_{cc}^{2D}=1-\left(\frac{b}{a}\right)^2$.  The median of $E_{cc}^{2D}$ according to the bins of mass and the bins of $\log\sigma_{NL}^{-1}$ are represented in \cref{tFoF2d}. The invariance is again found, the measurements on all the simulations, whatever the cosmological model, of this shape parameter as a function of the (tridimensional) non-linear fluctuations of the cosmic field are very precisely superimposed. Such a cosmological invariance allows, again, to reconstruct the power spectrum of the 3D cosmic field.

\section{Accuracy of shape parameters measurements of low mass DM halos}
\label{appb}

There are two possible types of uncertainty in measuring the shape parameters of DM halos. The first one is statistical. It is determined by the number of halos from which the medians of the shape parameters are calculated. This statistical error is always negligible because there is always a very large number of halos in each catalog and for each mass bin. A second uncertainty in the measurement of the shape parameters is systematic. It is fixed by the number of particles on which the inertia tensor and therefore the shape parameters of each DM halo are calculated. This uncertainty is linked to the bounded or unbounded nature of the shape parameters of the DM halos and dominates at low masses, for which the number of particles making up the halo is smaller. To estimate the magnitude of this uncertainty, we calculated the shape parameters (ellipticity and prolaticity) and the axis ratios (b/a and c/a) for a set of particles distributed uniformly in a sphere. In principle, the shape parameters for such a distribution should be zero, and the axis ratios equal to 1. 

\begin{figure*}
    \centering
    \subfloat[Ellipticity versus number of particles for an spherical repartition ]{\includegraphics[width=\columnwidth]{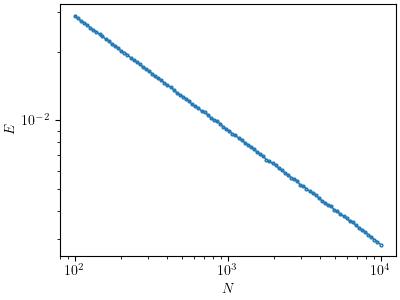}}
    \subfloat[Prolaticity versus number of particle]{\includegraphics[width=\columnwidth]{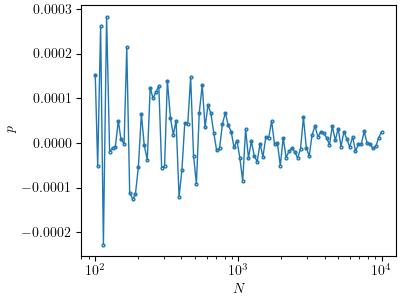}}\\
    \subfloat[$b/a$ axis ratio versus number of particles]{\includegraphics[width=\columnwidth]{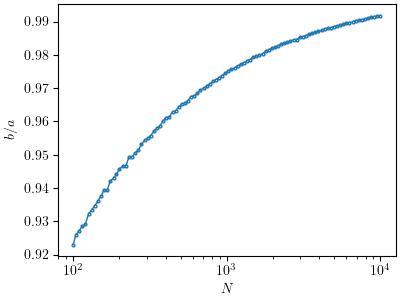}}
    \subfloat[$c/a$ axis ratio versus number of particles]{\includegraphics[width=\columnwidth]{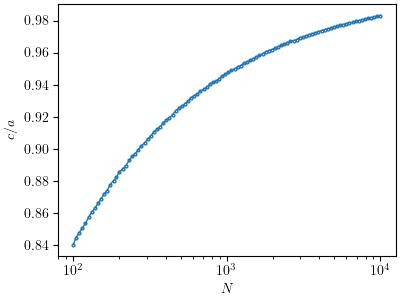}}
    \caption{Shape parameters (ellipticity and prolaticity of a set of particles uniformly distributed in a sphere. These two parameters have to be equal to $0$. In practice, the ellipticity decreases with the number of particles, the prolaticity tends towards $0$ when the number of particles tends towards infinity. The axis ratios ($b/a$ and $c:a$) tend towards $1$ when the number of particles increases. The accuracy of the ellipticity measurement is of the order of $10^{-2}$, which is sufficient to correct the ellipticity measurement observed in figures 14d and 16d to reach the value corresponding to cosmological universality. The accuracy of the measurement of the c/a ratio is sufficient to explain the discrepancy observed in Figures 14c and 16c and thus again re-establish cosmological universality.  }
    \label{figB1}
\end{figure*}

We present in figures \ref{figB1}, the measurements of these quantities of a set of $N$ particles varying between $100$ and $10000$ distributed in a sphere, for each value of $N$ we make the average of 10000 independent measurements. Since ellipticity is bounded below zero, it is always, at finite $N$, strictly greater than zero. The average of $E$ calculated over the 10,000 repetitions at $N$ is therefore systematically positive. For halos of 1000 particles (corresponding, in our simulation, to halos of mass $2\cdot 10^{12}$), the ellipticity is estimated in particular, on average, at $10^{-2}$ (instead of 0), i.e. of the same order of magnitude as the deviation from universality observed for these halos in figures 14d and 16d.

Prolaticity, on the other hand, is always less than $10^{-4}$ (positively or negatively), which is very low compared with the typical prolaticity of halos of mass $10^{12} M_\odot/h$ observed in Figures 13 and 15. This ensures that the prolaticity is insensitive to this systematic bias. The same applies to triaxiality, which is of course bounded in the segment [0,1], but whose typical values for halos of any mass are very far from the values of these bounds and which we have not reproduced here to lighten this appendix.The values of $b/a$ and $c/a$ ratio remain strictly below $1$. 

The shape measurements for low-mass halos close to the sphere are thus biased towards an overestimation of ellipticity and an underestimation of axis ratios. Such uncertainties are sufficient for low-mass halos, made up of a smaller number of particles, to correct the discrepancies observed in the ellipticity and $c/a$ ratio in figures 14c, 14d and 16c, in order to remain in agreement with the universal cosmological invariance highlighted in this work. 

\end{document}